\documentclass[twocolumn]{aastex631}
\usepackage{mathtools}
\usepackage{dsfont}
\usepackage[nolist]{acronym} 
\usepackage[nameinlink,capitalise,noabbrev]{cleveref}

\defcitealias{moon21}{Paper~I}
\defcitealias{moon22}{Paper~II}

\shorttitle{MHD Simulations of Nuclear Rings}
\shortauthors{Moon et al.}

\begin{document}

\title{Effects of Magnetic Fields on Gas Dynamics and Star Formation in Nuclear Rings}

\author[0000-0002-6302-0485]{Sanghyuk Moon}
\affiliation{Department of Physics \& Astronomy, Seoul National
  University, Seoul 08826, Republic of Korea}
\affiliation{Department of Astrophysical Sciences, Princeton University,
  Princeton, NJ 08544, USA}
\author[0000-0003-4625-229X]{Woong-Tae Kim}
\affiliation{Department of Physics \& Astronomy, Seoul National University, Seoul 08826, Republic of Korea}
\affiliation{SNU Astronomy Research Center, Seoul National University, Seoul 08826, Republic of Korea}
\author[0000-0003-2896-3725]{Chang-Goo Kim}
\affiliation{Department of Astrophysical Sciences, Princeton University,
  Princeton, NJ 08544, USA}
\author[0000-0002-0509-9113]{Eve C.\ Ostriker}
\affiliation{Department of Astrophysical Sciences, Princeton University,
  Princeton, NJ 08544, USA}
\email{sanghyuk.moon@princeton.edu, unitree@snu.ac.kr}
\email{cgkim@astro.princeton.edu, eco@astro.princeton.edu}

\begin{acronym}
    \acro{AGN}{active galactic nucleus}
    \acro{MHD}{magnetohydrodynamic}
    \acro{SFR}{star formation rate}
    \acro{SN}{supernova}
    \acro{FUV}{far ultraviolet}
    \acro{ISM}{interstellar medium}
    \acro{MJI}{magneto-Jeans instability}
    \acro{CMZ}{Central Molecular Zone}
    \acro{CND}{circumnuclear disk}
    \acro{CR}{cosmic ray}
    \acro{PRFM}{pressure-regulated, feedback-modulated}
\end{acronym}

\begin{abstract}
Nuclear rings at the centers of barred galaxies are known to be strongly magnetized.
To explore the effects of magnetic fields on star formation in these rings and nuclear gas flows, we run magnetohydrodynamic simulations in which there is a temporally-constant magnetized  inflow to the ring, representing a bar-driven inflow.
The mass inflow rate is $1\,M_\odot\,\mathrm{yr}^{-1}$, and we explore models with a range of field strength in the inflow.
We adopt the TIGRESS framework developed by Kim \& Ostriker to handle radiative heating and cooling, star formation, and resulting supernova (SN) feedback.
We find that magnetic fields are efficiently amplified in the ring due to rotational shear and SN feedback.
Within a few $100\,\mathrm{Myr}$, the turbulent component $B_\mathrm{trb}$ in the ring saturates at $\sim 35\,\mu\mathrm{G}$ (in rough equipartition with the turbulent kinetic energy density), while the regular component $B_\mathrm{reg}$ exceeds $50\,\mu\mathrm{G}$.
Expanding superbubbles created by clustered SN explosions vertically drag predominantly-toroidal fields from near the midplane to produce poloidal fields in high-altitude regions.
The growth of magnetic fields greatly suppresses star formation at late times.
Simultaneously, strong magnetic tension in the ring drives radially inward accretion flows from the ring to form a circumnuclear disk in the central region; this feature is absent in the unmagnetized model.
\end{abstract}


\section{Introduction}\label{s:intro}

A characteristic result of dynamical interactions between a bar and gas in disk galaxies is the formation of a pair of large-scale shocks running along the leading sides of the bar inside of corotation.
Gas entering the shock front loses angular momentum and is deflected inward.
In optical images, the compressed inflowing gas is seen as narrow dust lanes along which gas is funnelled toward the central regions.
The observed mass inflow rate is of the order of $\dot{M}_\mathrm{in} \sim 0.1$--$10\,\mathrm{M}_\odot\,\mathrm{yr}^{-1}$ and is believed to be time-variable \citep{benedict96, regan97, meier08, elmegreen09, shimizu19, sormani19}.

Bar-driven inflowing gas has residual angular momentum and thus forms a circumnuclear ring, which is often observed to be active in star formation.
Star-forming nuclear rings are found in about $\sim20\%$ of disk galaxies in the local universe, $80\%$ among which are barred \citep{comeron10}, and have \acp{SFR} of $\sim 0.1$--$10\,\mathrm{M}_\odot\,\mathrm{yr}^{-1}$ \citep{mazzuca08, ma18}.
While nuclear rings are sometimes found in unbarred galaxies, a majority of such galaxies have oval distortions, strong spiral arms, or close companions, all of which are thought to provide non-axisymmetric gravitational torques similar to bars that could effectively drive gas inward \citep{comeron10}.
Spectroscopic observations have revealed that nuclear rings are long-lived, composed of not only young star clusters formed recently but also old stellar populations with ages ranging from $\sim 100\,\mathrm{Myr}$ to a few $\mathrm{Gyrs}$.
The reconstructed star formation histories are characterized by a large time variability, involving multiple timescales ranging from a few tens of Myrs to a few Gyrs \citep[e.g.,][]{allard06, sarzi07, gadotti19, prieto19, nogueras-lara20}.
Over time, ring star formation may lead to the development of nuclear disks \citep{launhardt02,bittner20,gadotti20,freitas22,sormani22}, which are also known as ``disk-like bulges'' as distinct from classical and box/peanut bulges \citep{athanassoula05}.

Recently, a number of authors have  studied gas dynamics and star formation in and around nuclear rings, using numerical simulations with realistic treatment of star formation and feedback.
For example, \citet{armillotta19} conducted hydrodynamic simulations of the \ac{ISM} to study gas flows and star formation in the \ac{CMZ}, which is believed to represent a nuclear ring in our own Milky Way.
Their simulations (with a mass resolution of $2\times 10^3\,M_\odot$) showed that the \ac{SFR} of the \ac{CMZ} goes through several burst-quench cycles with a mixture of a short period ($\sim 50\,\mathrm{Myr}$) and long period ($\sim 200\,\mathrm{Myr}$), although the gas mass remains relatively constant over time.
\citet{tress20} and \citet{sormani20} used a higher mass resolution of $<100\,M_\odot$ (with adaptive mass refinement depending on local density and temperature) to model the \ac{CMZ}, resolving Sedov-Taylor blastwaves for most \ac{SN} explosions in their simulations.
Contrary to \citet{armillotta19}, these authors found that the \ac{SFR} in the ring steadily increases in time in proportion to the gas mass, with the gas depletion time almost constant within a factor of $\sim 2$.
\citet{wyseo19} ran hydrodynamic simulations coupled with $N$-body stellar dynamics to study how a nuclear ring forms and evolves in a situation where a stellar bar forms and grows self-consistently, rather than being treated as a fixed potential.
They found that star formation in a nuclear ring is sustained for a long ($>1\,\mathrm{Gyr}$) period of time, and that the ring \ac{SFR} correlates well with the mass inflow rate to the ring.

The diversity of findings from the above studies motivated us to undertake simulations with higher resolution in the ring region than it is possible to achieve with global models.
A key goal was to test whether a constant mass inflow rate results in steady ring star formation, or if instead the gas mass builds up and then produces intermittent bursts of star formation.
To explore whether steady vs.\ bursty behavior in ring star formation may depend on the inflow rate, in \citet{moon21} (hereafter \citetalias{moon21}) we developed a semi-global numerical framework that provides explicit control of the mass inflow rate via boundary conditions.
\citetalias{moon21} found that, (1) when the mass inflow rate is fixed in time, a quasi-steady equilibrium state is reached at $\text{SFR}\sim 0.8\dot{M}_\mathrm{in}$ for a wide range of $\dot{M}_\mathrm{in}$ ($0.125$--$8\,\mathrm{M}_\odot\,\mathrm{yr}^{-1}$), in which the \ac{SFR} and depletion time are almost constant within a factor of $\sim 2$; (2) vertical dynamical equilibrium is established within the ring gas, in which the thermal and turbulent pressures due to stellar feedback balance the gravitational field arising from both gas and stars; (3) the \ac{PRFM} star formation theory is satisfied as previously shown for featureless disks in \citet{cgkim17} and \citet{ostriker22}, and for disks with a spiral arm potential in \citet{wtkim20}, but in contrast to disk regions where the \ac{SFR} adapts to the equilibrium value set by gas mass, in nuclear rings the gas mass instead adapts to the \ac{SFR} set by the mass inflow rate.

To understand what might have produced temporal variations as well as spatial asymmetry of observed ring star formation, and how the delay between star formation and feedback affects interpretation of self-regulated equilibrium, in \citet{moon22} (hereafter \citetalias{moon22}) we allowed the mass inflow rate to vary with time in a prescribed way and/or to be spatially asymmetric.
\citetalias{moon22} found that (1) time-varying mass inflows with a sufficient oscillation amplitude cause episodic star formation, provided that the timescale of the inflow rate variations is sufficiently long ($\gtrsim 50\,\mathrm{Myr}$); (2) a sudden increase of the mass inflow rate through one of the two dust lanes causes lopsided star formation in the ring, lasting no longer than a few orbital period; (3) the \ac{PRFM} theory is still satisfied even for a non-steady system in which the \ac{SFR} and gas mass vary with time, provided that the time delay between star formation and feedback is properly taken into account.

While the studies above have improved our understanding of star-forming physics in nuclear rings, they were all limited to unmagnetized models.
Observations show that nuclear rings in real galaxies are quite strongly magnetized.
Assuming energy equipartition between magnetic fields and \acp{CR}, the average magnetic field strengths in nuclear rings inferred from radio synchrotron observations are estimated to be $\sim 55\,\mu\mathrm{G}$ for NGC 1097 \citep{beck05}, $\sim 63\,\mu\mathrm{G}$ for NGC 1365 \citep{beck05}, and $\sim 84\,\mu\mathrm{G}$ for NGC 5792 \citep{yang22}, much stronger than in spiral arms of normal disk galaxies \citep{beck15}.
\citet{beck99, beck05} mapped radio continuum emission in barred galaxies and found that the magnetic fields are predominantly parallel to the dust lanes, while penetrating the nuclear rings with a large pitch angle ($\sim 40^\circ$).
Strong magnetic fields would provide additional support for gas against gravity, potentially reducing the \ac{SFR} \citep{pillai15,tabatabaei18}.
Indeed, \citet{tabatabaei18} found a strong positive correlation between the gas depletion time and the magnetic field strength for individual giant clumps distributed along the nuclear ring of NGC 1097.
No correlation was found between the depletion time and the turbulent velocity dispersion, suggesting that it may be magnetic fields rather than \ac{SN} feedback that suppress ring star formation.

In this paper, we present results from \ac{MHD} simulations of star-forming, magnetized nuclear rings.
This work extends \citetalias{moon21} by considering magnetized gas inflows at the domain boundaries.
To focus on the effects of magnetic fields on ring star formation, we fix the mass inflow rate and characteristic ring radius (based on the imposed angular momentum of inflowing gas), while varying the magnetic field strength of the inflowing gas.
By comparing the results from models with different field strengths, we quantify how magnetic fields affect dynamical evolution of nuclear rings and star formation therein.

In addition to allowing us to study effects of magnetization on star formation, our models are useful to explore how magnetic fields affect accretion in the central region of galaxies.
In particular, our magnetized simulations show that gas accretes inward from the star-forming nuclear ring, which could potentially lead to formation of a \ac{CND} near the galactic center.
Based on the measured strength and pitch angle of magnetic fields in the nuclear ring of NGC 1097, \citet{beck05} suggested that magnetic stress can drive gas accretion from a ring to fuel an \ac{AGN}.
Other proposed mechanisms for gas inflows near a galaxy center include bars-within-bars \citep{shlosman89}, nuclear spirals \citep{maciejewski04, wtkim17}, and \ac{SN} feedback \citep{wada04, tress20}.
Here we use direct numerical simulations to study gas accretion and its outcomes in the presence of magnetic fields and star formation feedback.

The remainder of this paper is organized as follows.
In \autoref{s:method}, we outline the equations that we solve, summarize the TIGRESS\footnote{TIGRESS is an  acronym for ``Three-phase Interstellar medium in Galaxies Resolving Evolution with Star formation and Supernova feedback''} numerical framework for the \ac{ISM} and star formation physics, and describe our treatment of the boundary conditions for magnetized gas inflows.
In \autoref{s:evolution}, we present the overall time evolution of our models with a focus on star formation histories, and examine gas accretion toward the center driven by magnetic stresses.
In \autoref{s:bfield}, we present the temporal evolution of the magnetic field strength in the ring and explore the effects of magnetization on the ring star formation.
Finally, we summarize and discuss our results in \autoref{s:summary-discussion}.

\section{Numerical Methods}\label{s:method}

\begin{deluxetable}{lccccc}
    \tablecaption{Model parameters\label{tb:models}}
    \tablehead{
\colhead{Model} &
\colhead{$R_{\rm ring}$} &
\colhead{$\dot{M}_{\rm in}$} &
\colhead{$\beta_\mathrm{in}$} &
\colhead{${B}_{\mathrm{in},c}$} &
\colhead{$B_\mathrm{in,avg}$} \\
\colhead{(1)} &
\colhead{(2)} &
\colhead{(3)} &
\colhead{(4)} &
\colhead{(5)} &
\colhead{(6)} \\
\colhead{} &
\colhead{$({\rm pc})$} &
\colhead{$(M_\odot\,{\rm yr^{-1}})$} &
\colhead{} &
\colhead{$(\mu\mathrm{G})$} &
\colhead{$(\mu\mathrm{G})$}
    }
    \startdata
{\tt\ Binf} & 500 & 1.0 & $\infty$ & 0 & 0 \\
{\tt\ B100} & 500 & 1.0 & $100$ & 1.6 & 0.76\\
{\tt\ B30}  & 500 & 1.0 & $30$ & 3.0 & 1.4\\
{\tt\ B10}  & 500 & 1.0 & $10$ & 5.2 & 2.4
    \enddata
\end{deluxetable}

To numerically model the central kiloparsec region of a barred galaxy with high resolution, we adopt the semi-global numerical model introduced by \citetalias{moon21}.
In this approach, our computational domain covers only the nuclear ring and its immediate vicinity, and bar-driven mass inflows are treated by boundary conditions.
Nonlinear interactions between the bar and gas leading to the gas inflows are assumed to occur outside of the computational domain, and are not explicitly modeled.
Instead, we control the mass inflow rate and angular momentum of the inflows using free parameters.
In this section, we present the basic equations we solve (\autoref{s:governing-equations}), summarize the TIGRESS framework for star formation and feedback (\autoref{s:sf-feedback}), and describe the inflow boundary conditions for magnetized gas (\autoref{s:nozzle}).

\subsection{Governing Equations}\label{s:governing-equations}

Our computational domain is a Cartesian cube with side $L=2048\,\mathrm{pc}$ located at the galaxy center.
The domain rotates at an angular frequency $\mathbf{\Omega}_p = 36\,\mathrm{km\,s^{-1}\,kpc^{-1}}\hat{\mathbf{z}}$, corresponding to the adopted bar pattern speed\footnote{Even though we do not include a bar potential explicitly, it is advantageous to work in the rotating frame, since then the nozzles for inflow streams  (see section \ref{s:nozzle}) may be kept fixed in both space and time.}.
We include radiative heating and cooling of the \ac{ISM}, gaseous self-gravity, and a fixed external gravitational potential responsible for the background rotation curve.
The governing equations we solve are
\begin{equation}\label{eq:cont}
    \frac{\partial\rho}{\partial t} + \boldsymbol\nabla\cdot\left(\rho {\bf v}\right) = 0,
\end{equation}
\begin{equation}\label{eq:momentum}
\begin{split}
    \frac{\partial(\rho \mathbf{v})}{\partial t} + \boldsymbol\nabla\cdot\left(\rho \mathbf{v}\mathbf{v} + P\mathds{I}+\mathds{T}\right) \\= - 2\rho\mathbf{\Omega}_p\times\mathbf{v}-\rho\boldsymbol\nabla\Phi_\mathrm{tot},
\end{split}
\end{equation}
\begin{equation}\label{eq:energy}
\begin{split}
    \frac{\partial E}{\partial t} + \boldsymbol\nabla\cdot\left[E\mathbf{v} + (P\mathds{I}+\mathds{T})\cdot\mathbf{v}\right] \\= -\rho\mathbf{v}\cdot\boldsymbol\nabla\Phi_\mathrm{tot} - \rho\mathcal{L},
\end{split}
\end{equation}
\begin{equation}\label{eq:induction}
    \frac{\partial\mathbf{B}}{\partial t} = \boldsymbol\nabla\times(\mathbf{v}\times\mathbf{B}),
\end{equation}
\begin{equation}\label{eq:Poisson}
    \boldsymbol\nabla^2\Phi_\text{self} = 4\pi G(\rho + \rho_{\rm sp}).
\end{equation}
Here, $\rho$ and $\rho_\mathrm{sp}$ are respectively the volume density of gas and young star particles that form, $\mathbf{v}$ is the gas velocity in the rotating frame, $P$ is the gas pressure, $\mathds{I}$ is the identity matrix, $\mathds{T} = B^2/(8\pi)\mathds{I} - \mathbf{B}\mathbf{B}/(4\pi)$ is the Maxwell stress tensor, $E=\rho v^2/2 + P/(\gamma - 1) + B^2/(8\pi)$ is the total energy density with  adiabatic index $\gamma=5/3$, $\Phi_\mathrm{tot}=\Phi_\mathrm{self}+\Phi_\mathrm{ext}+\Phi_\mathrm{cen}$ is the total gravitational potential, consisting of the self-gravitational potential $\Phi_\mathrm{self}$, the external gravitational potential $\Phi_\mathrm{ext}$, and the centrifugal potential $\Phi_\mathrm{cen}=-\tfrac{1}{2}\Omega_p^2(x^2+y^2)$, and $\rho\mathcal{L}$ is the net cooling rate per unit volume.

In our models, the external gravity $\Phi_\mathrm{ext} = \Phi_\mathrm{BH} + \Phi_b$ arises from a central supermassive black hole and a stellar bulge.
We note that we do not include a nonaxisymmtric bar potential.
The black hole is modeled by a Plummer potential
\begin{equation}\label{eq:plummer}
    \Phi_\mathrm{BH} = -\frac{GM_\mathrm{BH}}{\sqrt{r^2+r_\mathrm{BH}^2}}
\end{equation}
with the mass $M_\mathrm{BH} = 1.4 \times 10^8\,M_\odot$ and the softening length $r_\mathrm{BH}=20\,\mathrm{pc}$.
For the stellar bulge, we take
\begin{equation}\label{eq:bulge-potential}
    \Phi_b = -\frac{4\pi G\rho_{b0}r_b^3}{r}\ln\left(\frac{r}{r_b}+\sqrt{1+\frac{r^2}{r_b^2}}\right).
\end{equation}
with the central density $\rho_{b0} = 50\,M_\odot\,\mathrm{pc}^{-3}$ and the scale radius $r_b = 250\,\mathrm{pc}$.
The resulting  rotation curve and the circular velocity at the ring position are similar to those of NGC 1097 as reported in \citet{onishi15}.

The net cooling rate of gas per unit volume in \cref{eq:energy} is given by
\begin{equation}\label{eq:net_cooling}
    \rho\mathcal{L} = n_\mathrm{H}(n_\mathrm{H}\Lambda - \Gamma_\mathrm{PE} - \Gamma_\mathrm{CR}),
\end{equation}
where $n_\mathrm{H}=\rho/(\mu_\mathrm{H}m_\mathrm{H})$ is the hydrogen number density with the mean molecular weight per hydrogen $\mu_\mathrm{H}=1.4271$ assuming the solar abundances.
For the cooling function $\Lambda(T)$, we take the fitting formula of \citet{koyama02} (see \citealt{cgkim08} for a typo-corrected version) for $T<10^{4.2}\mathrm{K}$, and the tabulated collisional ionization equilibrium cooling curve at solar metalicity of \citet{sutherland93} for $T>10^{4.2}\mathrm{K}$.
The gas temperature $T$ is related to density and pressure via an ideal equation of state $P = \rho k_\mathrm{B} T/(\mu m_\mathrm{H})$, with the mean molecular weight $\mu(T)$ varying with $T$ from $\mu_\mathrm{ato}=1.295$ for neutral gas to $\mu_\mathrm{ion}=0.618$ for fully ionized gas \citep{cgkim17}.

In \cref{eq:net_cooling}, $\Gamma_\mathrm{PE}$ represents an idealized model of the photoelectric heating rate by far-ultraviolet (FUV) radiation impinging on dust grains, and is given by
\begin{equation}\label{eq:gamma-pe}
    \Gamma_\mathrm{PE} = \Gamma_\mathrm{PE,0}\left(\frac{\mu(T) - \mu_\mathrm{ion}}{\mu_\mathrm{ato} - \mu_\mathrm{ion}}\right)\left(\frac{J_\mathrm{FUV}}{J_\mathrm{FUV,0}}+0.0024\right),
\end{equation}
where $\Gamma_\mathrm{PE,0} = 2\times 10^{26}\,\mathrm{erg\,s^{-1}}$ \citep{koyama02} and $J_\mathrm{FUV,0} = 2.1 \times 10^4\,\mathrm{erg\,s^{-1}cm^{-2}sr^{-1}}$ \citep{draine78} are normalization factors based on the solar neighborhood conditions.
The term in the first parentheses in \cref{eq:gamma-pe} reduces
the photoelectric heating  at high $T$ (since realistically dust grains would sublimate), shutting this heating off completely in the fully ionized gas.
The small factor in the last parentheses represents a minor contribution from the metagalactic \ac{FUV} background.

We use the same approximate method as in \citetalias{moon21} to calculate the mean FUV intensity $J_\mathrm{FUV}$ from young star particles in the simulations.
For this, we first calculate the luminosity surface density $\Sigma_\mathrm{FUV}$ of all star particles younger than $40\,\mathrm{Myr}$ in the simulation domain\footnote{The FUV luminosity-to-mass ratio of the star particles depends on their age based on \texttt{STARBURST99} model calculations.}, and then apply an approximate model for  dust attenuation, setting
\begin{equation}\label{eq:JFUV}
    J_\mathrm{FUV} = \frac{\Sigma_\mathrm{FUV}}{4\pi}\left(\frac{1-E_2(\tau_\perp/2)}{\tau_\perp}\right)e^{-n_\mathrm{H}/n_0}.
\end{equation}
Here, $E_2$ is the second exponential integral, $\tau_\perp=\kappa_d\Sigma$ with $\kappa_d=10^3\,\mathrm{cm^{-2}\,g^{-1}}$ is the vertical optical depth for the mean gas surface density $\Sigma$ averaged over the entire domain, and the factor in parentheses represents the average attenuation factor for a uniform-density slab with a uniform source distribution.
The exponential factor represents local shielding, with $n_0$ the density above which this shielding becomes significant.
For the models presented in this paper, we take $n_0=50\,\mathrm{cm^{-3}}$, which yields a dependence of $J_\mathrm{FUV}$ on density comparable to that obtained by applying the adaptive ray-tracing method of \citet{jgkim17} \citepalias[for details of this comparison see][]{moon21}.

Inside dense regions where \ac{FUV} radiation is heavily shielded, the heating is dominated by the \ac{CR} ionization.
The adopted heating rate $\Gamma_\mathrm{CR}$ in \cref{eq:net_cooling} is given by
\begin{equation}\label{eq:gamma-cr}
    \Gamma_\mathrm{CR} = q_\mathrm{CR}\xi_\mathrm{CR}\left(\frac{\mu(T) - \mu_\mathrm{ion}}{\mu_\mathrm{ato} - \mu_\mathrm{ion}}\right),
\end{equation}
where $q_\mathrm{CR}=10\,\mathrm{eV}$ is the energy yield per ionization \citep[see also \citealt{gong17}]{glassgold12} and $\xi_\mathrm{CR}$ denotes the \ac{CR} ionization rate.
The term inside the parentheses is again to shut off the CR heating in fully ionized gas.
Assuming that $\xi_\mathrm{CR}$ is proportional to the \ac{SFR} surface density $\Sigma_\mathrm{SFR}$ and is attenuated by a factor of $\Sigma_0/\Sigma$ above a critical gas surface density $\Sigma_0=10.7\,M_\odot\,\mathrm{pc}^{-2}$ \citep{neufeld17}, we set
\begin{equation}\label{eq:xi}
    \xi_\mathrm{CR} = \xi_\mathrm{CR,0}\frac{\Sigma_\mathrm{SFR}}{\Sigma_\mathrm{SFR,0}}\min\left\{1,\frac{\Sigma_0}{\Sigma}\right\},
\end{equation}
where $\xi_\mathrm{CR,0}=2\times10^{-16}\mathrm{s^{-1}}$ is the \ac{CR} ionization rate in the solar neighborhood \citep{indriolo07, neufeld17}.

\Crefrange{eq:cont}{eq:induction} are discretized on a uniform mesh with $512^3$ cells: the corresponding grid spacing is  $\Delta x=4\,\mathrm{pc}$.
We update the physical quantities using a version of the \texttt{Athena} MHD code \citep{stone08}, which employs the MUSCL-Hancock scheme with the constrained transport algorithm to preserve $\boldsymbol\nabla\cdot\mathbf{B}=0$ within machine precision \citep{stone09}, and applies the first-order flux correction when needed \citep{lemaster09}.
We apply the Green's function convolution method aided by a fast Fourier transform \citep[e.g.,][]{skinner15} to solve
the Poisson equation (Equation \ref{eq:Poisson}) with the vacuum boundary condition, i.e., $\Phi_\mathrm{self}\to 0$ at infinity.

\subsection{Star Formation and Feedback}\label{s:sf-feedback}

We handle star formation and feedback using the TIGRESS framework \citep[see also \citetalias{moon21}]{cgkim17}, which we briefly summarize here.
We refer the reader to \citet{cgkim17} for a more complete description.

We create a sink particle whenever the following three conditions are met simultaneously: (1) $\rho>\rho_\mathrm{LP}=8.86c_s^2/(\pi G\Delta x^2)$ with a local sound speed $c_s$, the threshold density based on the Larson-Penston collapse solution, (2) $\Phi_\mathrm{self}$ is a local minimum\footnote{$\Phi_\text{ext}$ and $\Phi_\text{cen}$ vary very slowly in space and thus have negligible contribution to the gradients of $\Phi_\text{tot}$.} and (3) the velocity is converging in all directions.
A portion of the gas mass in the surrounding 27 cells is converted to the initial mass of a newly created sink particle.
Sink particles are allowed to accrete mass and momentum from their surroundings and merge with nearby particles within $3\Delta x$ until the onset of the first \ac{SN} explosion ($\sim 4\,\mathrm{Myr}$).
For orbits of sink particles, we solve their equations of motion
\begin{equation}
\ddot{\mathbf{x}} = -\boldsymbol\nabla\Phi_\mathrm{tot} - 2\mathbf{\Omega}_p \times \dot{\mathbf{x}},
\end{equation}
using the Boris algorithm that preserves the Jacobi integral very accurately \citep[see also the Appendix of \citetalias{moon21}]{boris70}.

Sink particles with age less than $40\,\mathrm{Myr}$ exert feedback in the form of the photoelectric heating (\cref{eq:gamma-pe}), \ac{CR} heating (\cref{eq:gamma-cr}), and type II \ac{SN} explosions.
The method of energy and momentum injection for a given SN event depends on the density of the ambient medium.
If the ambient density is low enough that the shell-formation radius is expected to be resolved, we regard the \ac{SN} remnants as being in the Sedov-Taylor phase and inject $72\%$ of the \ac{SN} energy $E_\mathrm{SN}=10^{51}\,\mathrm{erg}$ in the form of thermal energy and the remaining $28\%$ in the form of kinetic energy.
If the ambient density is too high for the adiabatic stage of evolution to be resolved, we assume that the  \ac{SN} remnant has already cooled to enter the snowplow phase and inject radial momentum $p_*=2.8\times 10^5\,M_\odot\,\mathrm{km\,s^{-1}}(n_\mathrm{H}/\mathrm{cm}^{-3})^{-0.17}$ as calibrated from higher-resolution simulations \citep{cgkim15-SN}.
In both cases, each \ac{SN} event returns the ejecta mass $M_\mathrm{ej}=10\,M_\odot$ from a sink particle back to the \ac{ISM}.

\subsection{Magnetized Inflow Streams}\label{s:nozzle}

\begin{figure*}
    \plotone{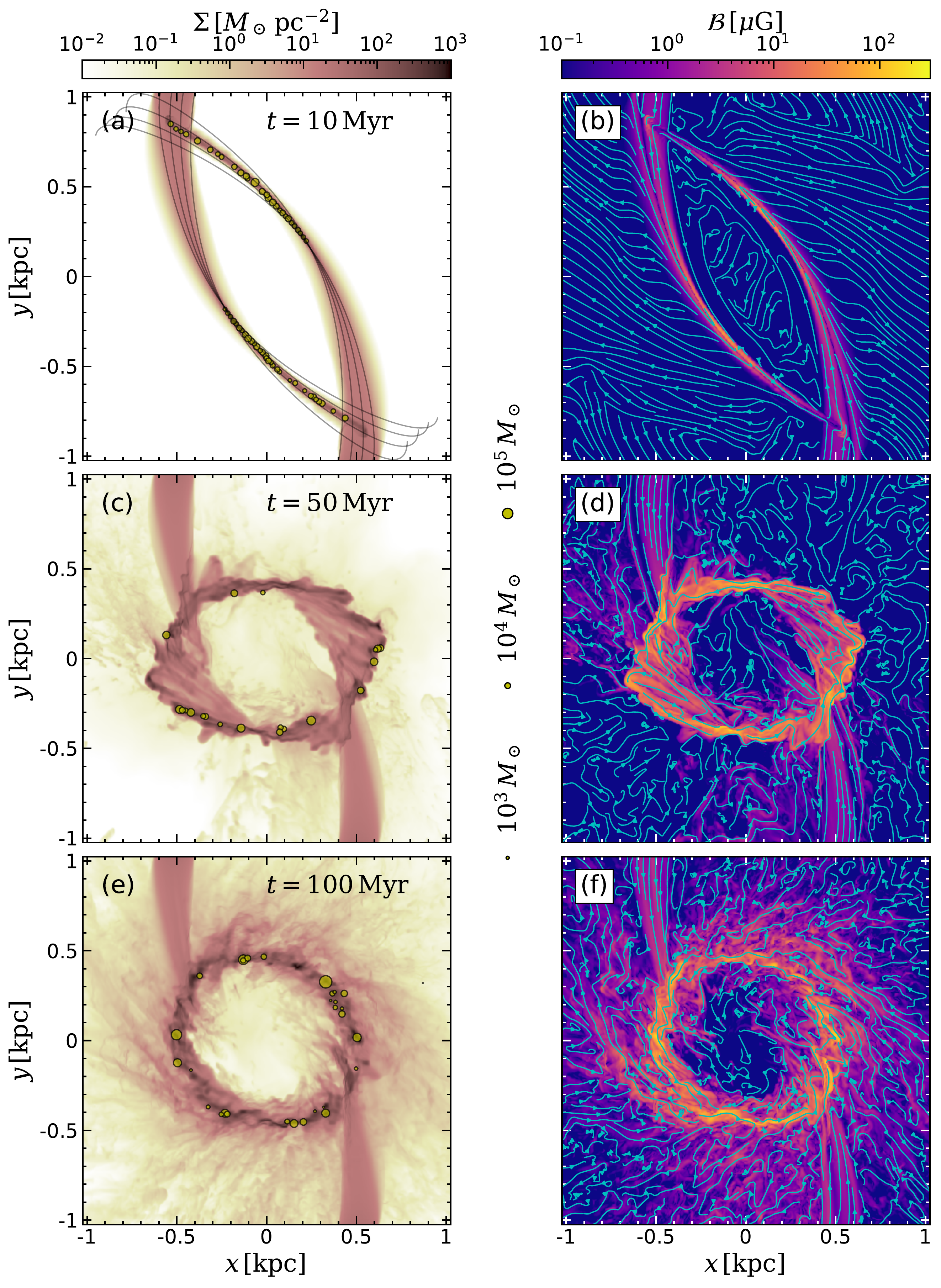}
    \caption{Face-on views of model \texttt{B100} at $t=10$, $50$, $100$, $220$, $250$, and $285\,\mathrm{Myr}$ (the figure continues on the next page). The left column displays the gas surface density (color scale) and newly formed star particles with age $<1\,\mathrm{Myr}$ (circles).  The right column plots streamlines of the projected magnetic fields $\mathcal{B}_x = (\int \rho B_x\,dz) / (\int \rho\,dz)$ and $\mathcal{B}_y = (\int \rho B_y\,dz) / (\int \rho\,dz)$, overlaid on maps of $\mathcal{B} = (\mathcal{B}_x^2 + \mathcal{B}_y^2)^{1/2}$ (color scale). The black solid lines in the panel (a) are the ballistic trajectories that a test particle injected with $\mathbf{v}_\mathrm{in}$ would follow.}
    \label{fig:evolution}
\end{figure*}

\begin{figure*}
    \addtocounter{figure}{-1}
    \plotone{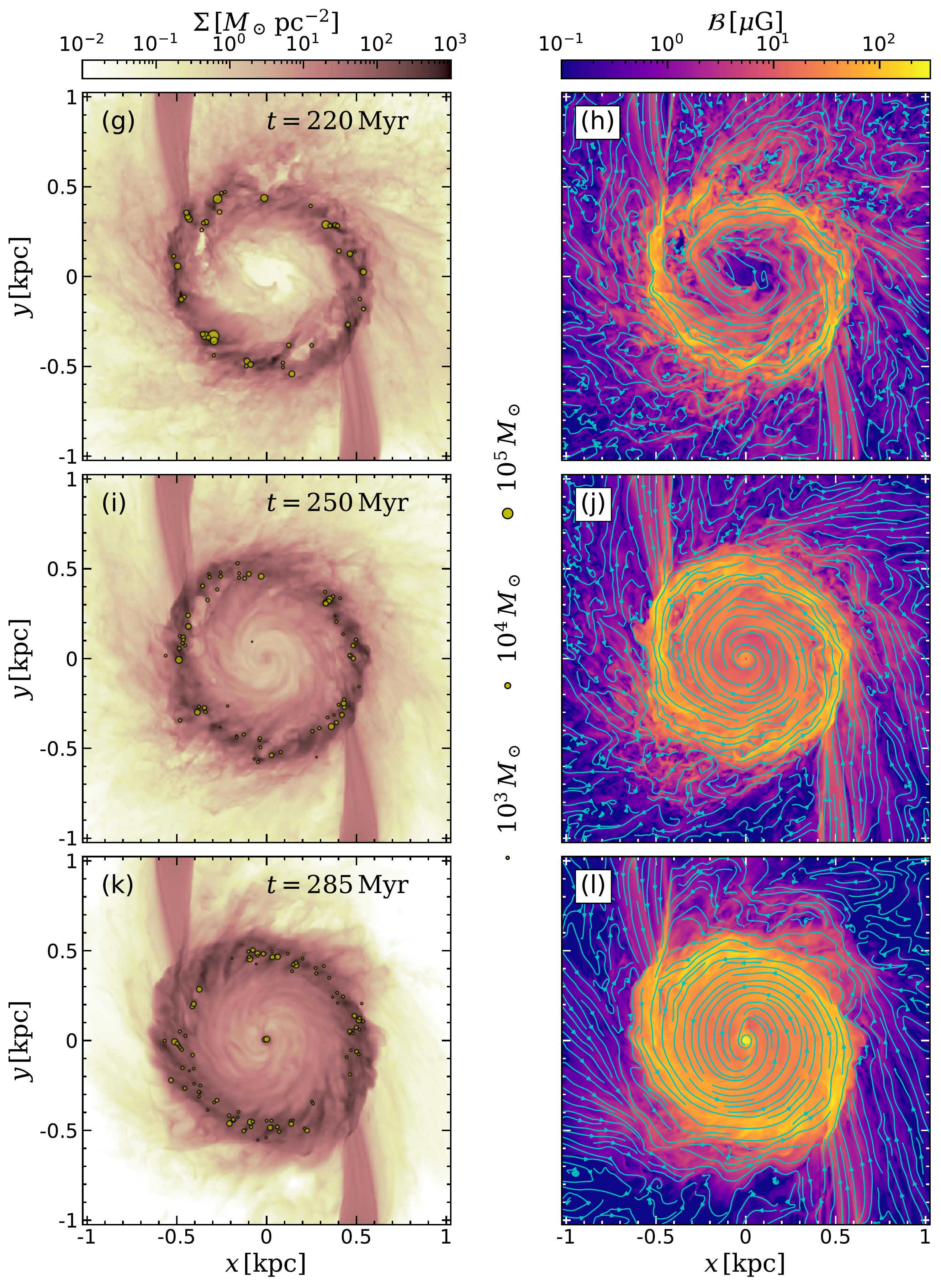}
    \caption{continued}
\end{figure*}

\citetalias{moon21} introduced the semi-global framework that treats bar-driven mass inflows via imposed boundary conditions for hydrodynamic simulations.
Here, we modify the boundary conditions slightly to handle magnetized inflows.

We inject gas streams into the computational domain through two circular nozzles with radius $\zeta_\mathrm{in}=112\,\mathrm{pc}$ placed at the $y$-boundaries: the coordinates of the nozzle centers are $(x,y,z)=(\mp b_\mathrm{in},\pm L/2, 0)$, where $b_\mathrm{in}=512\,\mathrm{pc}$ is the impact parameter of the inflows \citepalias[see Figure 3 of][for schematic diagram]{moon21}.
Here and hereafter, the upper and lower signs correspond to the upper and lower nozzles, respectively.
We set the streaming velocity at the nozzles to
\begin{equation}\label{eq:vin}
    \mathbf{v}_\mathrm{in} = \mp v_\mathrm{in}(\sin\theta_\mathrm{in}\hat{\mathbf{x}} + \cos\theta_\mathrm{in}\hat{\mathbf{y}}),
\end{equation}
where $\theta_\mathrm{in} = 10^\circ$ is the inclination angle of the streams relative to the $y$-axis.
The condition of the angular momentum conservation implies that the inflow speed $v_\mathrm{in}$ determines the location where the nuclear ring forms.
By setting the specific angular momentum (in the inertial frame) of the inflows equal to $R_\mathrm{ring}v_\mathrm{rot}(R_\mathrm{ring})$ with the circular velocity $v_\mathrm{rot}\equiv (Rd\Phi_\mathrm{ext}/dR)^{1/2}$, we obtain
\begin{equation}\label{eq:vin-mag}
    v_\mathrm{in}(x, \pm L/2) = \frac{R_\mathrm{ring}v_\mathrm{rot}(R_\mathrm{ring})-R^2\Omega_p}{|x\cos\theta_\mathrm{in} \mp (L/2)\sin\theta_\mathrm{in}|}.
\end{equation}
We fix the ring radius to $R_\mathrm{ring}=500\,\mathrm{pc}$ and use \cref{eq:vin,eq:vin-mag} to find the corresponding inflow velocity inside the nozzles, which varies from $72\,\mathrm{km\,s^{-1}}$ to $115\,\mathrm{km\,s^{-1}}$.

The density $\rho_\mathrm{in}$  of the streams sets the mass inflow rate as
\begin{equation}
    \dot{M}_\mathrm{in} = \iint \rho_\mathrm{in}v_\mathrm{in}\cos\theta_\mathrm{in}dxdz,
\end{equation}
where the integrations are performed over the two nozzles, i.e., $y=\pm L/2$ and $[(x\pm b_\mathrm{in})^2+z^2]^{1/2}< \zeta_\mathrm{in}$.
We fix the mass inflow rate to $1\,M_\odot\,\mathrm{yr}^{-1}$ by taking $\rho_\mathrm{in}=0.138\,M_\odot\,\mathrm{pc}^{-3}$, corresponding to $n_\mathrm{H} = 3.9\,\mathrm{cm}^{-3}$.
The mean inflow speed in the nozzles amounts to $\bar{v}_\mathrm{in} \equiv \dot{M}_\mathrm{in}/(2\rho_\mathrm{in}\pi \zeta_\mathrm{in}^2\cos\theta_\mathrm{in}) = 91\,\mathrm{km\,s^{-1}}$.

Radio polarization observations indicate that the magnetic fields are roughly parallel to dust lanes and point to the galactic center \citep{beck05,lopez-rodriguez21}.
Motivated by this, we take the magnetic fields inside the nozzles parallel to the inflow velocity as
\begin{equation}\label{eq:Bin}
    \mathbf{B}_\mathrm{in} = \frac{B_\mathrm{in}}{v_\mathrm{in}}\mathbf{v}_\mathrm{in},
\end{equation}
with the amplitude
\begin{equation}\label{eq:Bin-mag}
    B_\mathrm{in} = \left(\frac{8\pi P_\mathrm{in}}{\beta_\mathrm{in}}\right)^{1/2}\cos\left(\frac{\pi \zeta}{2\zeta_\mathrm{in}}\right).
\end{equation}
Here, $P_\mathrm{in} = \rho_\mathrm{in}k_\mathrm{B}T_\mathrm{in}/(\mu_\mathrm{H}m_\mathrm{H})$ is the thermal pressure of the inflowing gas with temperature $T_\mathrm{in}=2\times 10^4\,\mathrm{K}$, $\beta_\mathrm{in}$ is a plasma parameter measuring the ratio of thermal to magnetic pressure, and $\zeta = [(x\pm b_\mathrm{in})^2 + z^2]^{1/2}$ is the distance from the nozzle center.
The cosine term ensures that the fields vanish at the nozzle boundaries, preventing the gas just outside the nozzles from accidentally acquiring too large Alfv\'en speeds.
In \texttt{Athena}, the velocity and magnetic fields are cell-centered and face-centered, respectively.
Despite \cref{eq:Bin}, the mismatch in the evaluation points of  $\mathbf{B}_\mathrm{in}$ and  $\mathbf{v}_\mathrm{in}$ yields non-vanishing $\mathbf{v}_\mathrm{in}\times\mathbf{B}_\mathrm{in}$ at the innermost ghost zones at early time, as explained in \autoref{app:seed-magnetic-fields}.
This allows seed magnetic fields to leak into our computational domain through \cref{eq:induction}, which are subsequently stretched by the inflows to become parallel to the streams, smoothly matching the boundary conditions (see \autoref{s:overall-evolution}).

We allow gas to freely escape from the simulation domain, but forbid inflows except through the nozzles.
We accomplish this by setting the hydrodynamic variables in the ghost zones by extrapolating from the two adjacent active zones, while keeping the normal velocity to zero if the velocity is directed inward.
The magnetic fields in the ghost zones are simply copied from the innermost active zones.

\subsection{Models}

We consider four models with $\beta_\mathrm{in}=\infty$, $100$, $30$, and $10$ for the plasma beta parameter in the inflow.
\Cref{tb:models} summarizes the model parameters for all models.
Column (1) lists the model names.
Columns (2) and (3) give the ring radius and the mass inflow rate, respectively, which are the same for all models.
Column (4) gives $\beta_\mathrm{in}$.
Column (5) and (6) give the magnetic field strength at the nozzle centers $B_{\mathrm{in},c} = B_\mathrm{in}(\zeta=0)$ and the mean field strength inside the nozzles $B_\mathrm{in,avg}$, respectively.
Model \texttt{B100} is our fiducial model which has $\beta_\mathrm{in}=100$, $B_{\mathrm{in},c} = 1.6\,\mu\mathrm{G}$, and $B_\mathrm{in,avg} = 0.76\,\mu\mathrm{G}$.
The simulation domain is initially filled with rarefied gas with density $n_\mathrm{H}=10^{-5}\exp[-|z|/(50\,\mathrm{pc})]\,\mathrm{cm}^{-3}$ and temperature $T=2\times 10^4\,\mathrm{K}$, and subsequent evolution is  governed entirely by the inflowing streams.

\begin{figure*}
    \plotone{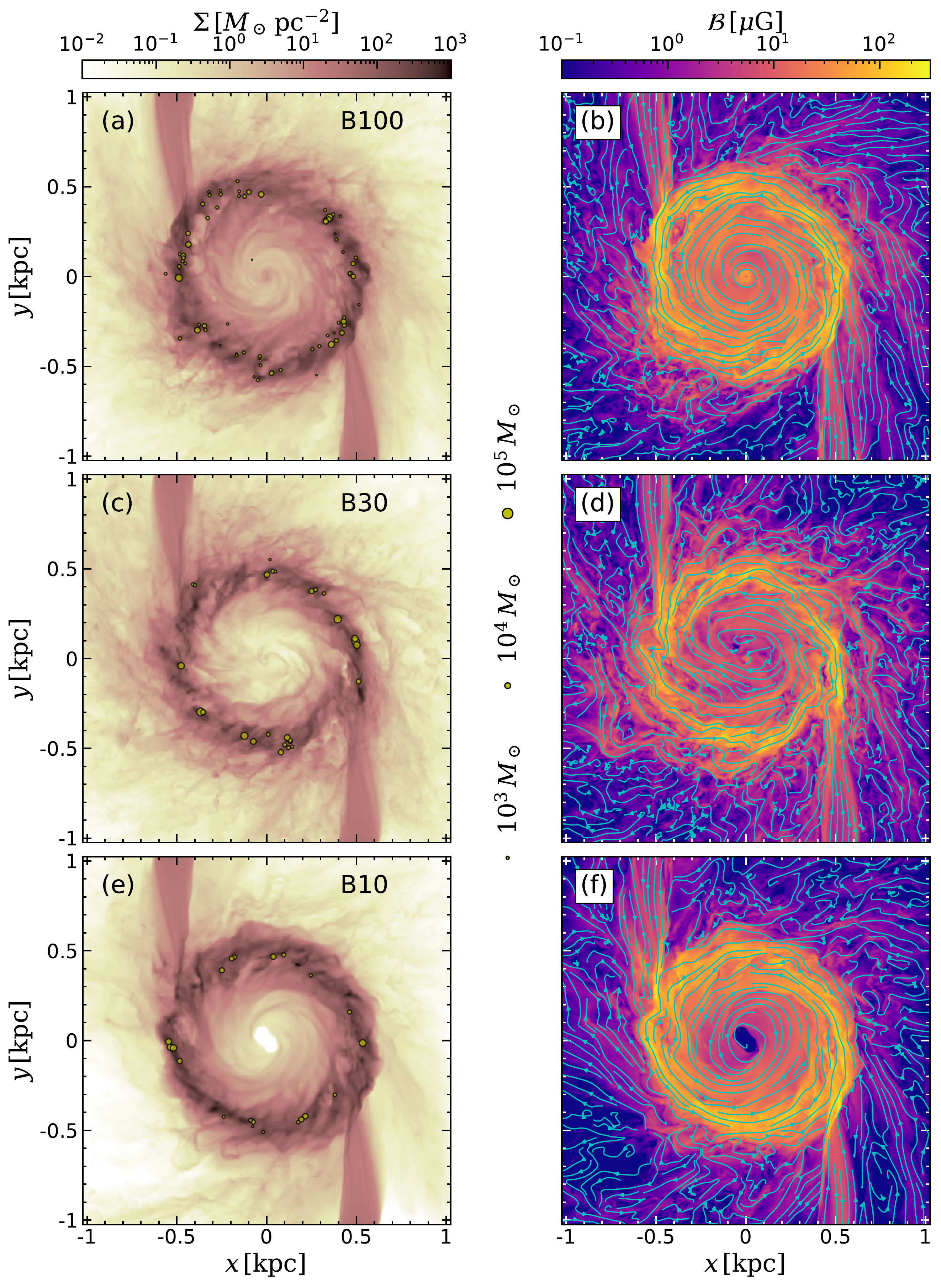}
    \caption{Similar to \cref{fig:evolution}, but for models \texttt{B100}, \texttt{B30}, and \texttt{B10} at $t=250$, $130$, and $130\,\mathrm{Myr}$, respectively (from top to bottom).}
    \label{fig:all_models}
\end{figure*}

\begin{figure*}
    \plotone{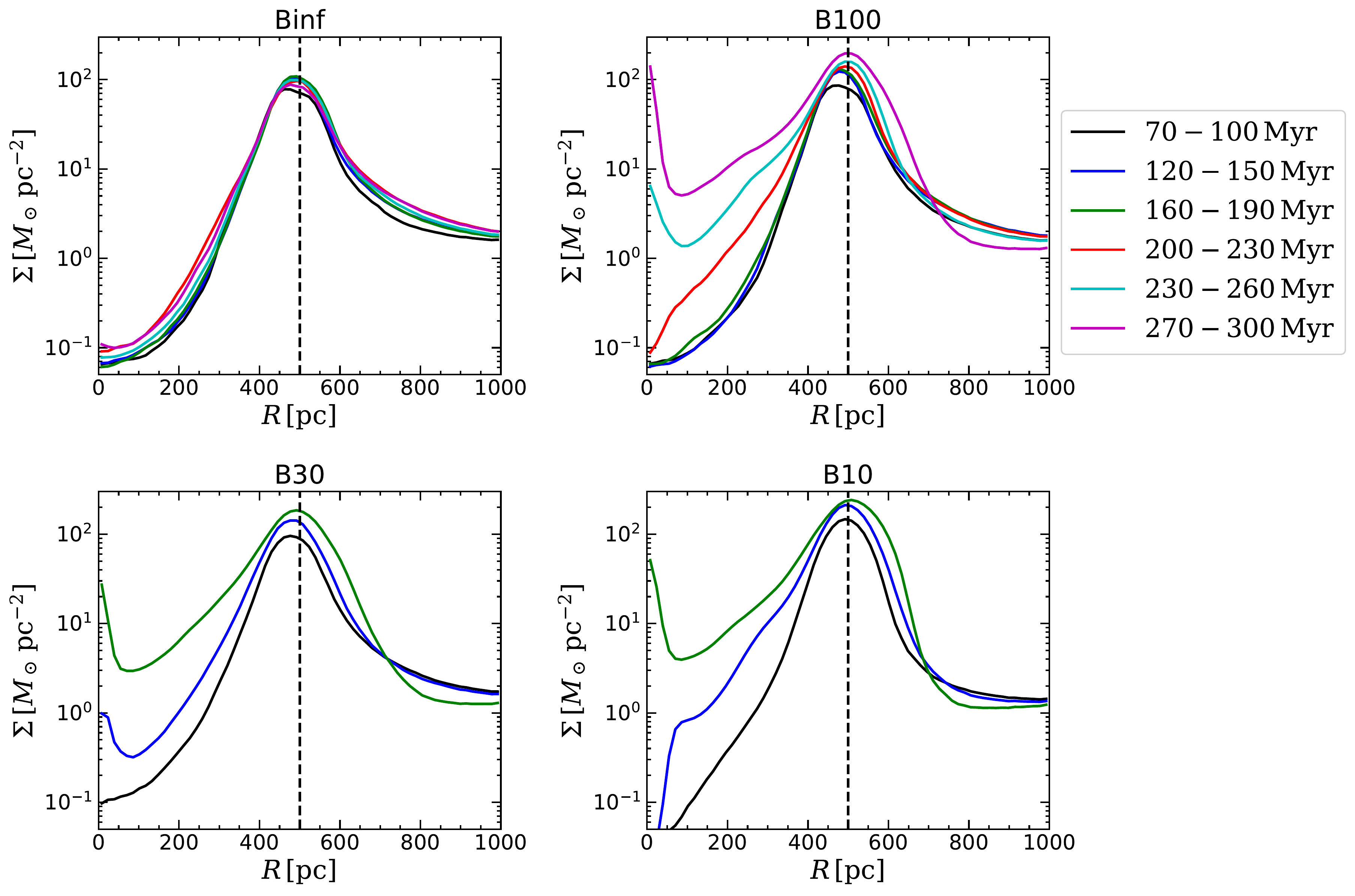}
    \caption{Evolution of the radial distribution of the azimuthally-averaged gas surface density for all models.
    Colors indicates the time interval for a temporal average. The vertical dashed lines mark the ring location $R_\mathrm{ring}=500\,\mathrm{pc}$.}
    \label{fig:surf_evolution}
\end{figure*}

\section{Evolution}\label{s:evolution}

\begin{figure*}
    \plotone{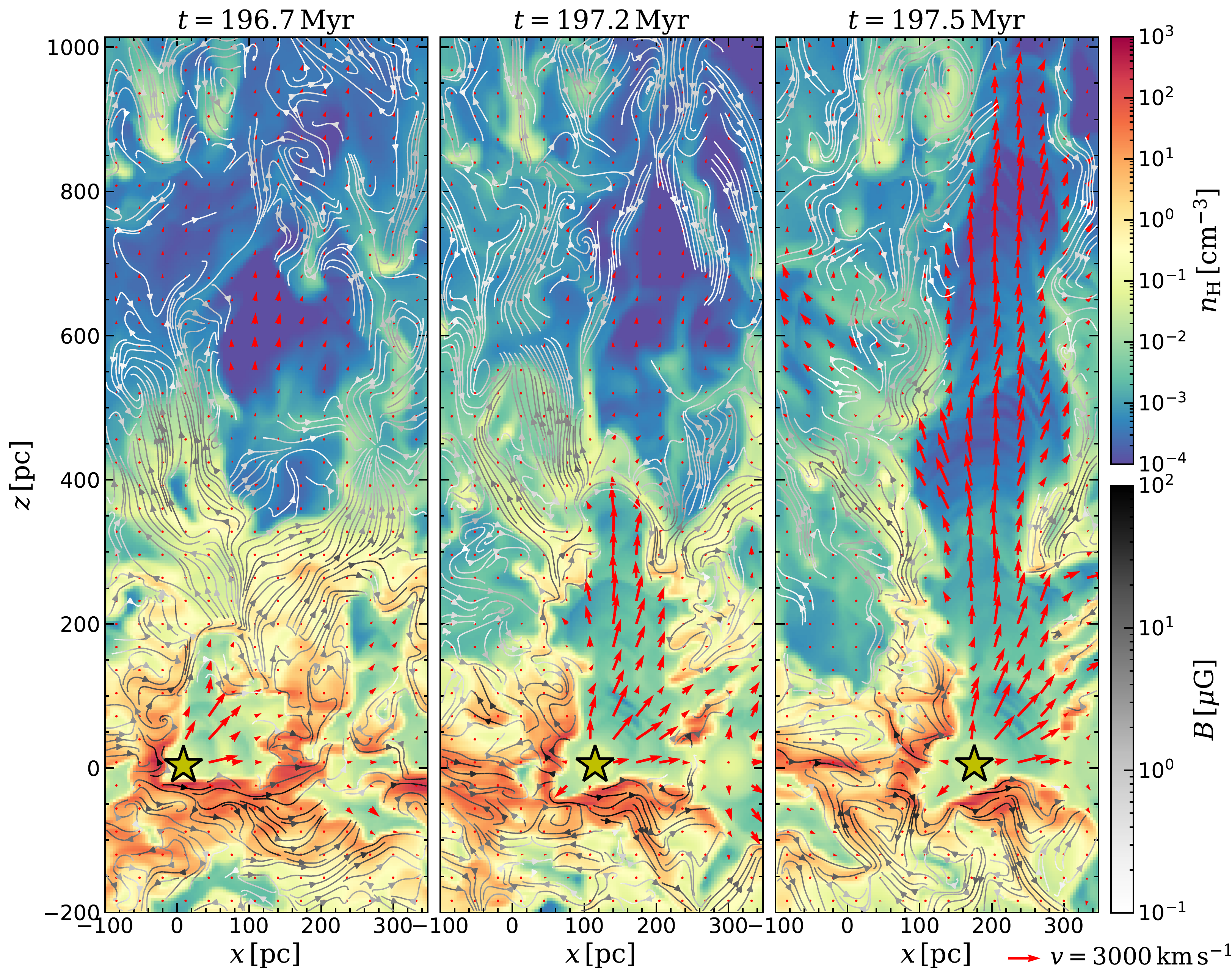}
    \caption{Superbubble breakout in the vertical direction. Each panel from left to right shows the density slice of model \texttt{B100} at $y=-438$, $-433$, and $-419\,\mathrm{pc}$, the $y$-position of a  sink particle with mass $2\times 10^6\,M_\odot$ (denoted by the yellow star) at $t=196.7$, $197.2$, and $197.5\,\mathrm{Myr}$, respectively. The streamlines in grey represent the magnetic fields lines with the strength $B=(B_x^2+B_y^2+B_z^2)^{1/2}>0.1\,\mu\mathrm{G}$. The red arrows are in-plane velocity vectors, $(v_x, v_z)$, with their lengths proportional to the speed $(v_x^2 + v_z^2)^{1/2}$. The expanding superbubble surrounding the star particle lifts the toroidal magnetic fields near the midplane to high-altitude regions to produce poloidal fields.
    }
    \label{fig:superbubble}
\end{figure*}

\begin{figure*}
    \plotone{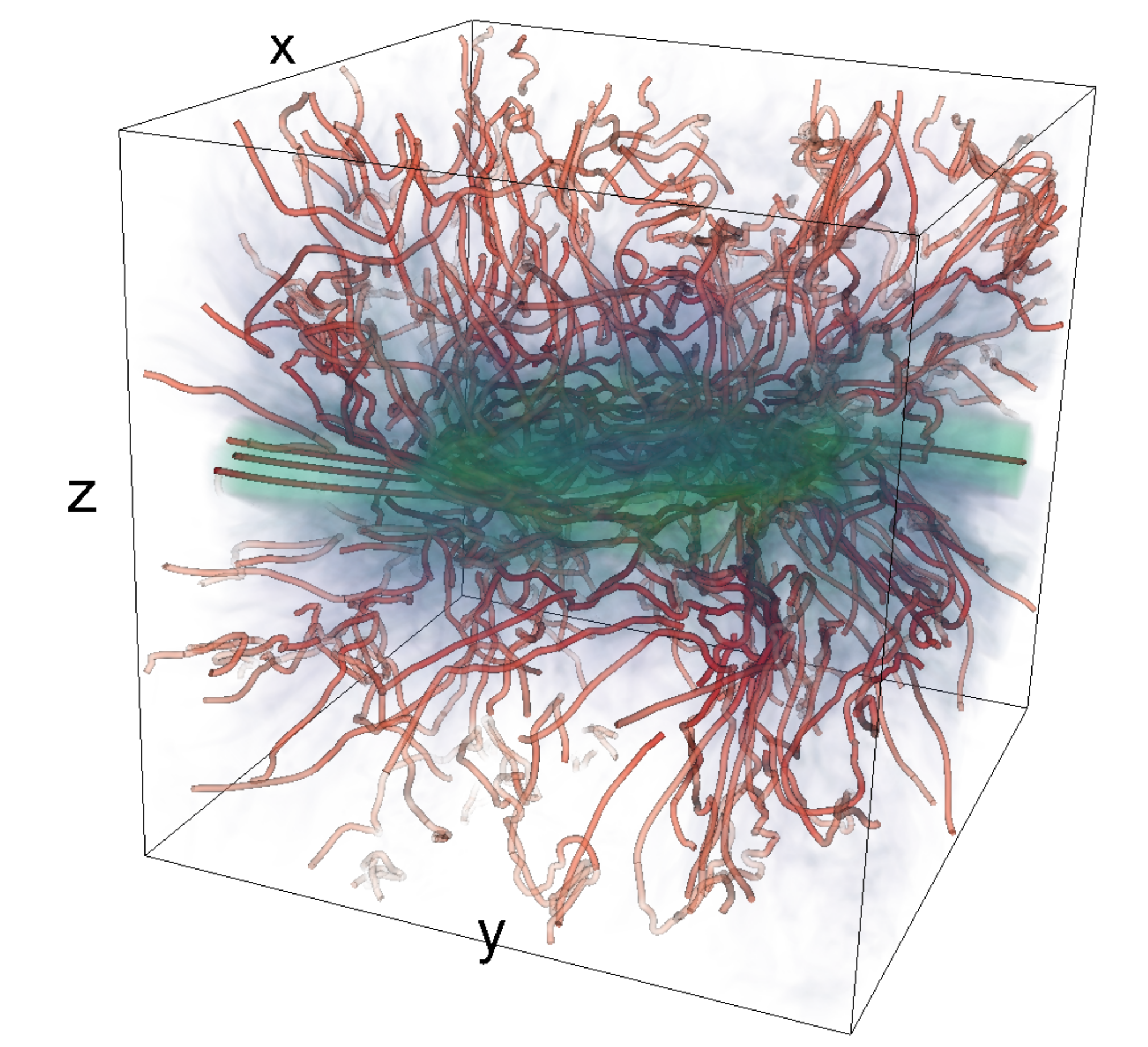}
    \caption{Perspective visualization of the three-dimensional magnetic field structure in model \texttt{B100} at $t=250\,\mathrm{Myr}$. The magnetic field lines are represented by red tubes, while the gas density is volume rendered in blue-green. Note that the magnetic fields are predominantly toroidal inside the nuclear ring because of the differential rotation, and poloidal in high-$|z|$ regions due to \ac{SN}-driven outflows.}
    \label{fig:volume-rendering}
\end{figure*}

In this section, we describe overall evolution of our fiducial model \texttt{B100} in terms of the gas and magnetic field distribution and star formation.
We also measure the accretion rates inside the ring and compare them with theoretical predictions.

\subsection{Overall Evolution}\label{s:overall-evolution}

\Cref{fig:evolution} plots snapshots of gas surface density together with young star particles as well as the projected magnetic field lines overlaid over the total field strength map in our fiducial model \texttt{B100} at a few selected epochs.
\Cref{fig:all_models} shows similar plots for models \texttt{B100}, \texttt{B30}, and \texttt{B10} at a selected epoch for each model.
\Cref{fig:surf_evolution} plots evolution of the radial profiles of the azimuthally-averaged surface density for all models.

Early evolution of model \texttt{B100} is qualitatively similar to that of the unmagnetized models presented in \citetalias{moon21}.
There is an initial transient phase during which the inflows follow nearly ballistic orbits (\cref{fig:evolution}(a)), but within half an orbital time ($\sim 8\,\mathrm{Myr}$), the streams from the opposite boundaries collide with each other, which drives strong shocks with a Mach number $\sim 16$.
The streams lose their orbital kinetic energy as they passes through the shocks multiple times, and form a nuclear ring with radius $R\sim R_\mathrm{ring}$ at $t\sim 50\,\mathrm{Myr}$ (corresponding to $\sim 3$ orbital times; see \cref{fig:evolution}(c)).

Still, the ring is elongated with the major axis precessing under the external gravitational potential.
It takes another $\sim 50\,\mathrm{Myr}$ for the ring to fully circularize (\cref{fig:evolution}(e,f)).
The ring soon reaches a quasi-steady equilibrium where the \ac{FUV} and \ac{CR} heating balances the radiative cooling, the \ac{SN} feedback balances the turbulent dissipation, and the thermal and turbulent pressures remain approximately constant.
The resulting total midplane pressure matches the overlying weight in the ring.
The gas mass in the ring also stays roughly constant as the net mass inflow rate balances the star formation rate.
Star formation proceeds randomly throughout the whole of the ring.
Although the resulting \ac{SN} feedback disperses the gas and drives turbulence locally and temporarily, it never destroys the ring entirely nor quenches star formation completely (\autoref{s:history}; see also \citetalias{moon21}).

As mentioned in \autoref{s:nozzle} (and \autoref{app:seed-magnetic-fields}), our boundary conditions introduce weak seed magnetic fields in the active domain, which are stretched along the streams by the inflowing gas.
Except for initial $\sim 10\,\mathrm{Myr}$, magnetic fields in the streams remain well aligned with the inflow velocity and do not exhibit systematic growth in time, although they are perturbed intermittently by strong \ac{SN} feedback from the ring.
As the streams form a ring, magnetic fields become predominantly toroidal in the ring, with large fluctuations due to \ac{SN} feedback.
The magnetic fields in the ring become stronger and more regular with time  (see \autoref{s:field-growth}), presumably due to both small- and large-scale dynamo driven by \ac{SN} feedback and rotational shear inside the ring, the discussion of which we defer to \autoref{s:dynamo}.

Strong magnetic fields in the ring cause evolution of model \texttt{B100} to deviate significantly from that of model \texttt{Binf} after $t\sim 200\,\mathrm{Myr}$ in two ways.
First, in contrast to model \texttt{Binf} where the surface density profile does not change much with time (\cref{fig:surf_evolution}), an accretion flow develops in model \texttt{B100} from the ring toward the center, gradually filling the region inside the ring.
The accreting gas piles up at the center, forming a \ac{CND} with radius of $\sim 50\,\mathrm{pc}$.
Second, strong magnetic fields and associated pressure make the \ac{SFR} decrease with time in model \texttt{B100} (see \autoref{s:effects-of-bfields-on-sf}).

\Cref{fig:evolution}(k) shows that much of the ring gas is concentrated in dense, trailing spiral segments with a pitch angle of $\lesssim 45^\circ$ and azimuthal spacing of $\sim 100$--$150\,\mathrm{pc}$.
These spiral segments start to appear roughly at $t\sim 260$--$270\,\mathrm{Myr}$ and keep being destroyed and regenerated thereafter.
The quasi-regular spacing of these spiral segments suggests that they result from the \ac{MJI} in which magnetic tension forces from bent field lines suppress the stabilizing effect of epicyclic motions \citep{elmegreen87,wtkim01,wtkim02}.
Indeed, the corresponding dispersion relation \citep[Equation 21 of][]{wtkim02} for the parameters adopted from model \texttt{B100} yields the most unstable wavelength of $\sim 120\,\mathrm{pc}$, entirely consistent with the numerical results.
Due to the \ac{MJI}, some spiral segments attain sufficient density to form stars.
We note, however, that strong shear in the ring makes the \ac{MJI} operate only temporarily, preventing runaway growth of spiral segments (see \autoref{s:effects-of-bfields-on-sf}).

\Cref{fig:superbubble} plots the spatial distribution of gas and magnetic fields in $x$--$z$ slices through three consecutive positions of a moving star cluster with mass $2\times 10^6\,M_\odot$, marked by the star symbol in each panel.
Repeated \ac{SN} explosions create a superbubble around the cluster.
The overpressurized bubble easily expands in the vertical direction where the gas density decreases, eventually breaking out and rapidly rising up with velocities exceeding $10^3\,\mathrm{km\,s^{-1}}$.
Magnetic field lines are lifted to high-$|z|$ regions and stretched by the flows of hot gas to generate a polodial component.
\Cref{fig:volume-rendering} plots a volumetric rendering of the three-dimensional magnetic field geometry and gas density, showing that the magnetic fields are predominantly toroidal in the ring and poloidal in the regions away from the midplane.

Evolution of models \texttt{B10} and \texttt{B30} is qualitatively similar to that of model \texttt{B100} in the sense that the inflow streams collide to form a star-forming nuclear ring, and accretion flows develop from the ring toward the center when the magnetic stress becomes strong enough (see \autoref{s:inflow}).
However, the models with smaller $\beta_\mathrm{in}$ reach the evolutionary stage characterized by decreased \ac{SFR} and an accretion flow from the ring toward the center at earlier time, compared to higher $\beta_\mathrm{in}$ models (see \cref{fig:all_models}).
\Cref{fig:surf_evolution} shows that unlike in model \texttt{Binf}, the rings in models \texttt{B100}, \texttt{B30}, and \texttt{B10} expand inward with time.
The surface density interior to the magnetized rings increases with time due to the radial accretion flows, forming a \ac{CND} characterized by the central upturn of the radial surface density profile at $R\lesssim 50\,\mathrm{pc}$.
We note that we are unable to evolve the magnetized models for arbitrarily long time, because the Alfv\'en speed becomes too large in the low-density region above and below the magnetized \ac{CND}, severely limiting the Courant-Friedrichs-Lewy timestep.

\subsection{Star Formation History}\label{s:history}

\begin{figure}
    \plotone{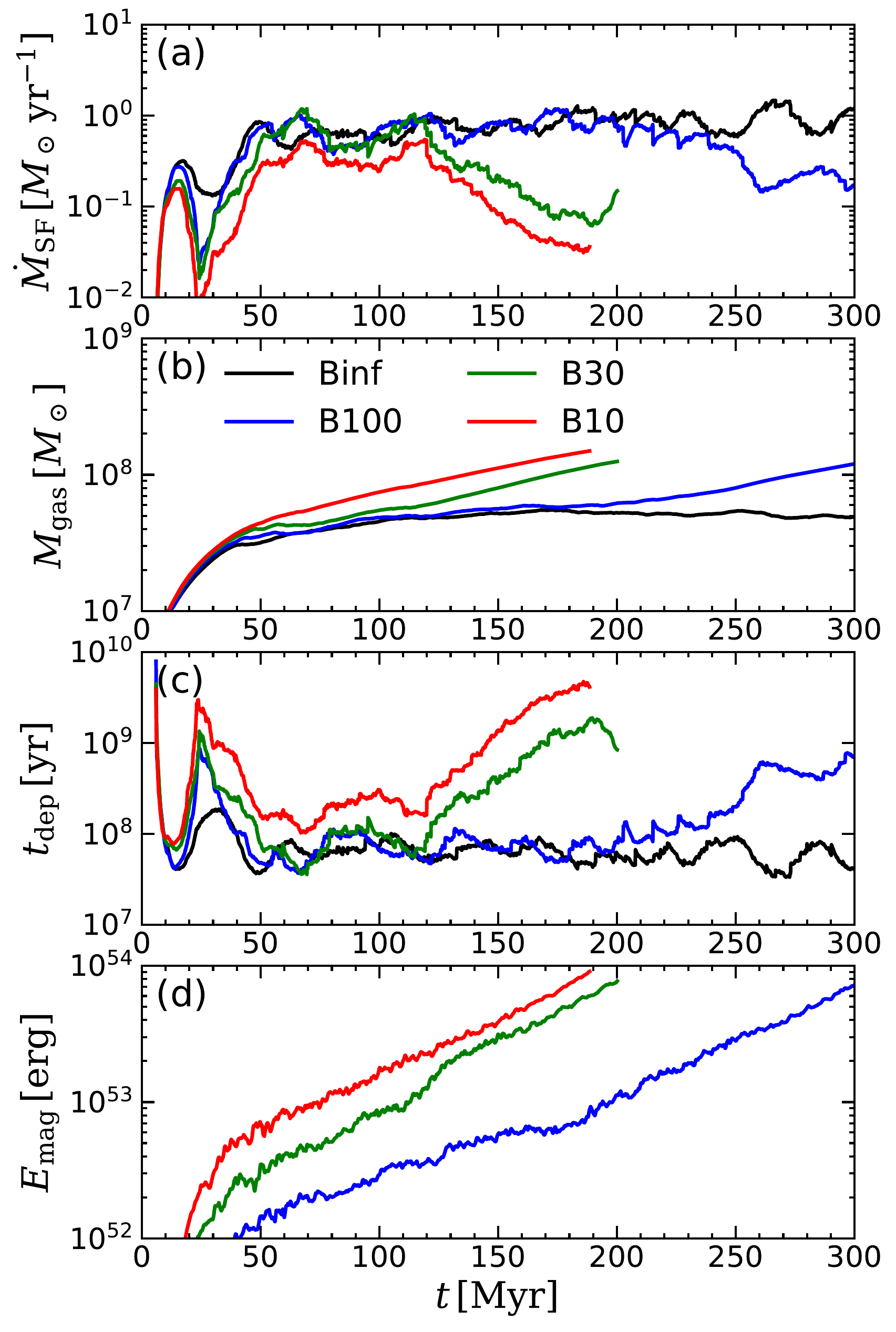}
    \caption{Temporal histories of (a) the \ac{SFR} $\dot{M}_\mathrm{SF}$, (b) the total gas mass $M_\mathrm{gas}$, (c) the gas depletion time $t_\mathrm{dep}$, and (d) the total magnetic energy inside the computational domain. The black, blue, green, and red lines correspond to models \texttt{Binf}, \texttt{B100}, \texttt{B30}, and \texttt{B10}, respectively.}
    \label{fig:history}
\end{figure}

We define the \ac{SFR}, $\dot{M}_\mathrm{SF}$, as the total mass of sink particles with age less than $10\,\mathrm{Myr}$, divided by $10\,\mathrm{Myr}$.
\Cref{fig:history} plots the temporal histories of $\dot{M}_\mathrm{SF}$, the total gas mass $M_\mathrm{gas}$, the gas depletion time
\begin{equation}
    t_\mathrm{dep}=\frac{M_\mathrm{gas}}{\dot{M}_\mathrm{SF}},
\end{equation}
and the total magnetic energy $E_\mathrm{mag}$ inside the computational domain.
For all models, there is initial transient behavior as the ring forms and star formation develops.
After $t\sim 50\,\mathrm{Myr}$, the \ac{SFR} becomes almost constant in model \texttt{Binf}, reaching a steady-state value $\dot{M}_\mathrm{SF}\sim 0.8$--$0.9\,M_\odot\,\mathrm{yr}^{-1}$, with a factor of $\sim 2$ stochastic fluctuations due to turbulence driven by \ac{SN} feedback, similar to the models presented in \citetalias{moon21}.

The star formation history of model \texttt{B100} is very similar to that of model \texttt{Binf} until $t\sim 200\,\mathrm{Myr}$.
After $t\sim 200\,\mathrm{Myr}$, however, strong magnetic fields in the ring of model \texttt{B100} result in a reduced \ac{SFR}.
The evolution of the \ac{SFR} in models \texttt{B30} and \texttt{B10} is qualitatively similar: it reaches a quasi-steady value at $t\sim 50$--$100\,\mathrm{Myr}$, which is lower by a factor of a few in the models with stronger magnetic fields (smaller $\beta_\mathrm{in}$),  and then starts to decline after $t\sim 120\,\mathrm{Myr}$.
The secular trend of increasing $M_\mathrm{gas}$ evident in \cref{fig:history}(b) as well as increasing gas surface density in the ring (\cref{fig:surf_evolution}) indicate that the decline of the \ac{SFR} in magnetized models is not caused by the reduction in the gas mass or surface density.
It is rather because a larger fraction of the ring gas becomes inert for star formation, as reflected in  \cref{fig:history}(c) which shows that $t_\mathrm{dep}$ increases at late time.
As the \ac{SFR} drops below $\dot{M}_\mathrm{in}$, the excess gas piles up in the ring and moves toward the center to form a \ac{CND}.

\Cref{fig:history}(d) shows that the magnetic energy in the computational domain exponentially increases with time, indicative of dynamo action.
We note that the magnetic energy advected with the inflow streams is very small because $\mathbf{v}\parallel\mathbf{B}$ near the nozzles for most of the time, and therefore cannot account for the increase of $E_\mathrm{mag}$ (see \autoref{app:magnetic-energy-conservation}).
We will present more detailed analysis on the growth of magnetic fields and their effects on star formation, and discuss possible causes of the magnetic field amplification in \autoref{s:bfield}.

\subsection{Magnetically Driven Accretion Flow}\label{s:inflow}

\begin{figure}
    \plotone{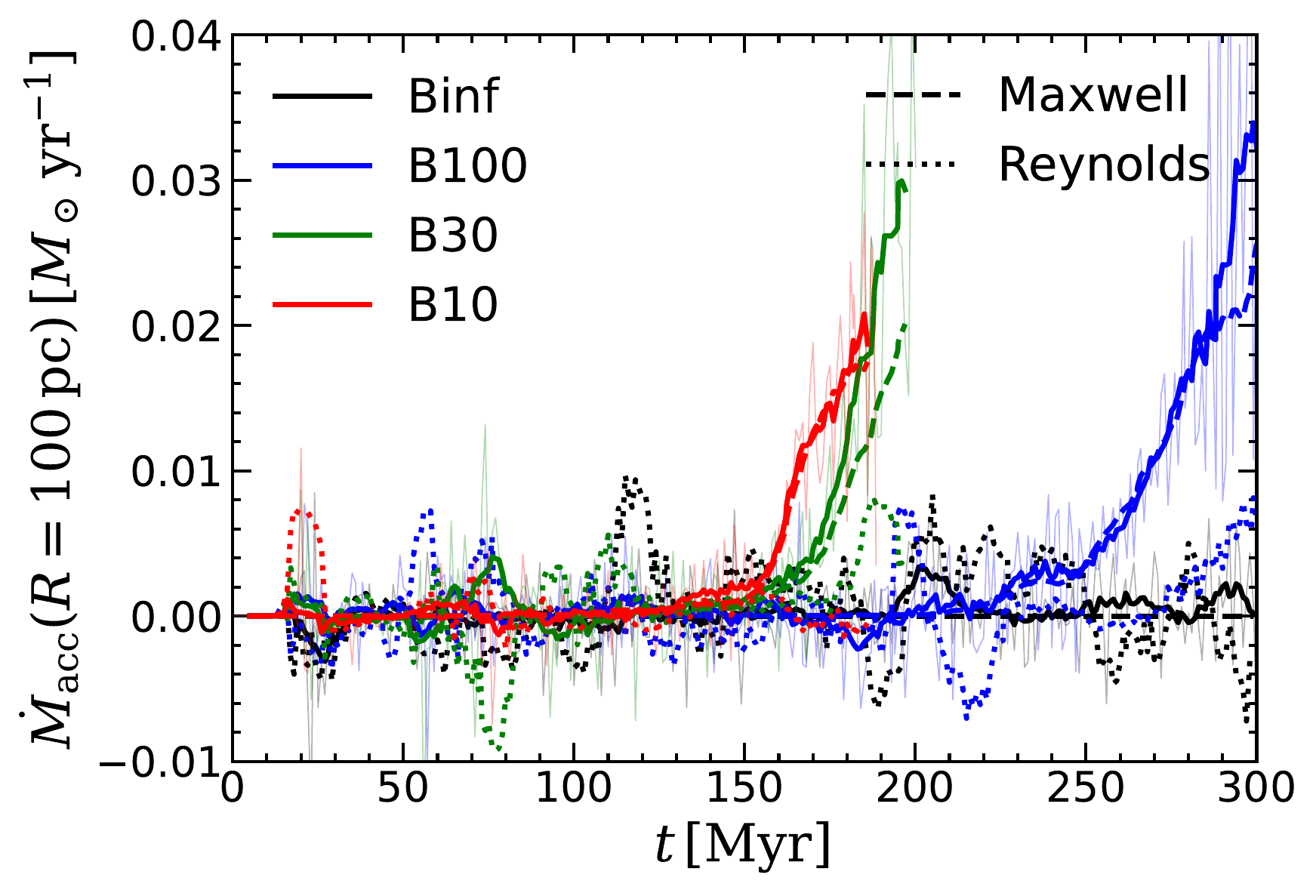}
    \caption{Temporal histories of the accretion rate $\dot{M}_\mathrm{acc}$ at $R=100\,\mathrm{pc}$ for models \texttt{Binf} (black), \texttt{B100} (blue), \texttt{B30} (green), and \texttt{B10} (red). Thin and thick solid lines correspond to the instantaneous and time-averaged (using a $10\,\mathrm{Myr}$ window) values, respectively, directly measured from the simulations. Dashed and dotted lines are the predicted accretion rates due to the Maxwell and Reynolds stresses, using \cref{eq:appmdot} (averaged over a $10\,\mathrm{Myr}$ window), respectively. The increasing trend of $\dot{M}_\mathrm{acc}$ in magnetized models is well explained by the Maxwell stress.}
    \label{fig:inflow_rate_t}
\end{figure}

\begin{figure}
    \plotone{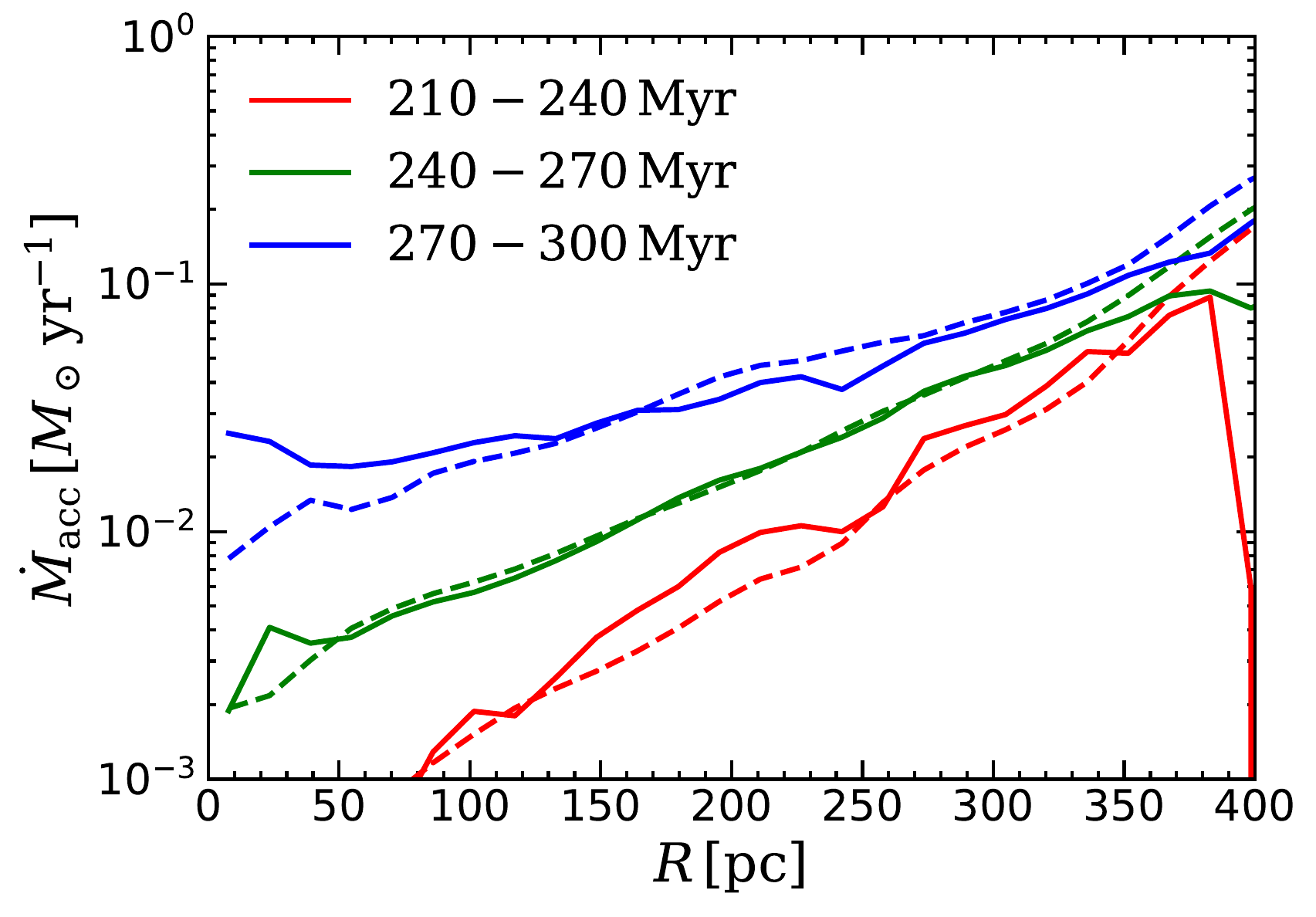}
    \caption{
    Radial profiles of the mass accretion rate at different epochs for model \texttt{B100}. Solid and dashed lines correspond to the measured accretion rate and predicted accretion rate due to the Maxwell stress, respectively.}
    \label{fig:inflow_rate_R}
\end{figure}

\Cref{fig:evolution,fig:surf_evolution} show that all  magnetized models develop an accretion flow from the ring toward the center.
For rotating magnetized disks,  \autoref{app:accretion} shows that the total mass accretion rate in a quasi-steady state can be written as
\begin{equation}
    \dot{M}_\mathrm{acc} \approx \dot{M}_\mathrm{M} + \dot{M}_\mathrm{R},
\end{equation}
where $\dot{M}_\mathrm{M}$ and $\dot{M}_\mathrm{R}$ are the mass accretion rates due to the Maxwell and Reynolds stresses, respectively, defined as
\begin{subequations}\label{eq:appmdot}
\begin{align}\label{eq:mdot-Maxwell}
    \dot{M}_\mathrm{M} &= 2\pi \left[\frac{\partial\left( R v_\mathrm{circ}\right)}{\partial R}\right]^{-1} \frac{\partial \left<R^2T_{R\phi}\right>}{\partial R}, \\
\label{eq:mdot-Reynolds}
    \dot{M}_\mathrm{R} &= 2\pi \left[\frac{\partial \left(R v_\mathrm{circ}\right)}{\partial R}\right]^{-1}\frac{\partial \left<R^2\rho u_R u_\phi\right>}{\partial R},
\end{align}
\end{subequations}
where $v_\mathrm{circ}=(\Omega - \Omega_p)R$ is the background circular velocity in the rotating frame and $(u_R,u_\phi)$ are the perturbation in the radial and azimuthal velocity.
\Cref{fig:inflow_rate_t} plots the temporal histories of $\dot{M}_\mathrm{acc}$ measured at $R=100\,\mathrm{pc}$ for all models, in comparison with the predictions due to the Reynolds and Maxwell stresses.
In model \texttt{Binf}, $\dot{M}_\mathrm{acc}\sim 4\times 10^{-4}\,M_\odot\,\mathrm{yr}^{-1}$ on average, showing that the mass accretion rate without magnetic fields remains small for all time.
While the values of $\dot{M}_\mathrm{acc}$ in the magnetized models are also small at early times, they increase rapidly to reach $\dot{M}_\mathrm{acc} \sim 0.02$--$0.03\,M_\odot\,\mathrm{yr}^{-1}$ toward the end of the runs.
The temporal changes of $\dot{M}_\mathrm{acc}$ are in good agreement with $\dot{M}_\mathrm{M}$, indicating that magnetic tension is the major driver of mass accretion in our models.

\Cref{fig:inflow_rate_R} plots the radial profiles of $\dot{M}_\mathrm{acc}$ averaged for a few selected time intervals for model \texttt{B100}, together with $\dot{M}_\mathrm{M}$.
The accretion rate is not constant in radius but decreases toward the center, with a slope decreasing with time.
This implies that mass is being deposited at all radii $<R_\mathrm{ring}$, consistent with the radial density distributions shown in  \cref{fig:surf_evolution}.
For $t=270$--$300\,\mathrm{Myr}$, the accretion rate near the ring is $\sim 0.1\,M_\odot\,\mathrm{yr}^{-1}$, i.e., one tenth of the bar-driven inflow rate, while it decreases to $0.02$--$0.03\,M_\odot\,\mathrm{yr}^{-1}$ near the center.
The radial dependence of the measured accretion rates is overall in good agreement with $\dot{M}_\mathrm{M}$, indicating that the accretion flows are mediated mostly by the magnetic tension forces.

\section{Magnetic Fields in the Ring}\label{s:bfield}

In this section, we analyze evolution of the regular and turbulent magnetic fields in nuclear rings and explore their effects on the ring star formation.
We also discuss vertical dynamical equilibrium in the presence of magnetic fields.
Finally, we discuss our results in the context of dynamo theory.

\subsection{Growth of Magnetic Fields}\label{s:field-growth}

\begin{figure*}
    \plotone{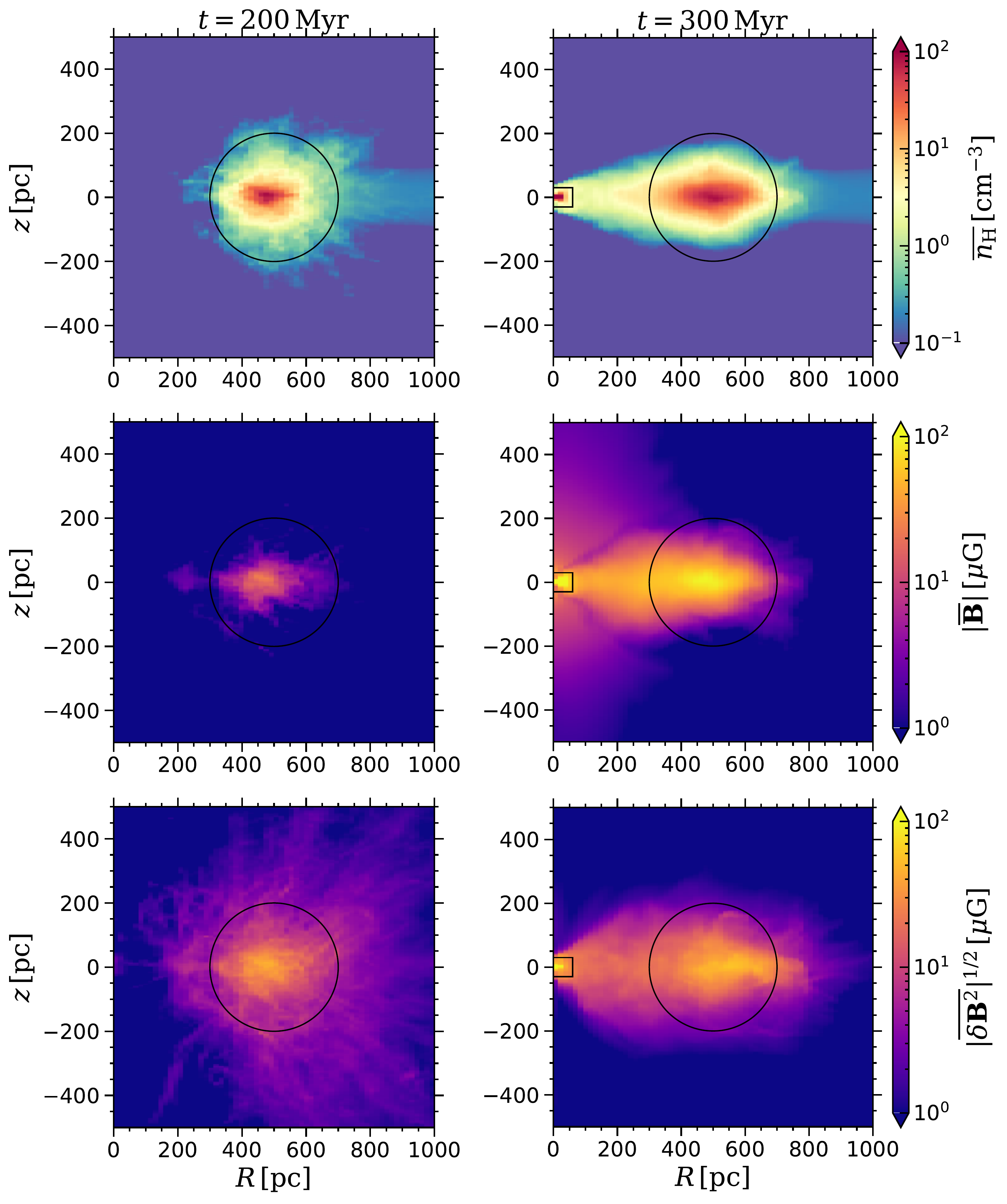}
    \caption{Spatial distributions from model \texttt{B100} of the azimuthally-averaged hydrogen number density (top), and the strength of the regular (middle) and turbulent (bottom) components of the magnetic fields at $t=200$ (left column) and $300\,\mathrm{Myr}$ (right column). The black circles centered at $(R,z)=(500, 0)\,\mathrm{pc}$ with radius $200\,\mathrm{pc}$ outline the ring, while the rectangles near $R=0$ in the right column mark a \ac{CND}.}
    \label{fig:rzplane}
\end{figure*}

\begin{figure}
    \plotone{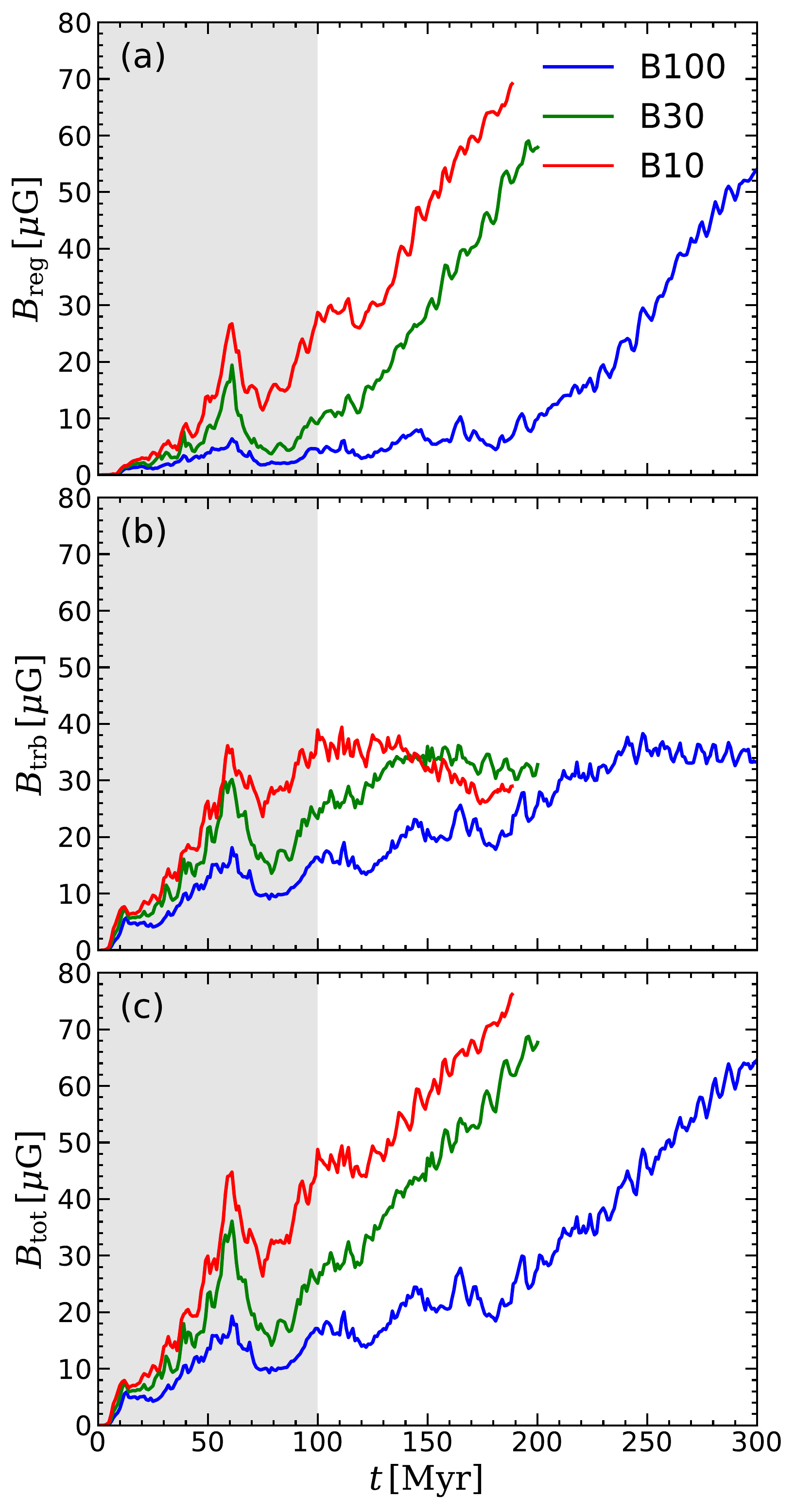}
    \caption{
    Temporal histories of the strength of the (a) regular, (b) turbulent, and (c) total magnetic field in the ring for models \texttt{B100} (blue), \texttt{B30} (green), and \texttt{B10} (red). Magnetic fields are  dominated by the turbulent component at early time, but become predominantly regular at late time. In the shaded regions at $t<100\,\mathrm{Myr}$, the rings are not circular so that the decomposition of the magnetic fields into the regular and turbulent components is not meaningful.
    }
    \label{fig:bfield_history}
\end{figure}

\begin{figure}
    \plotone{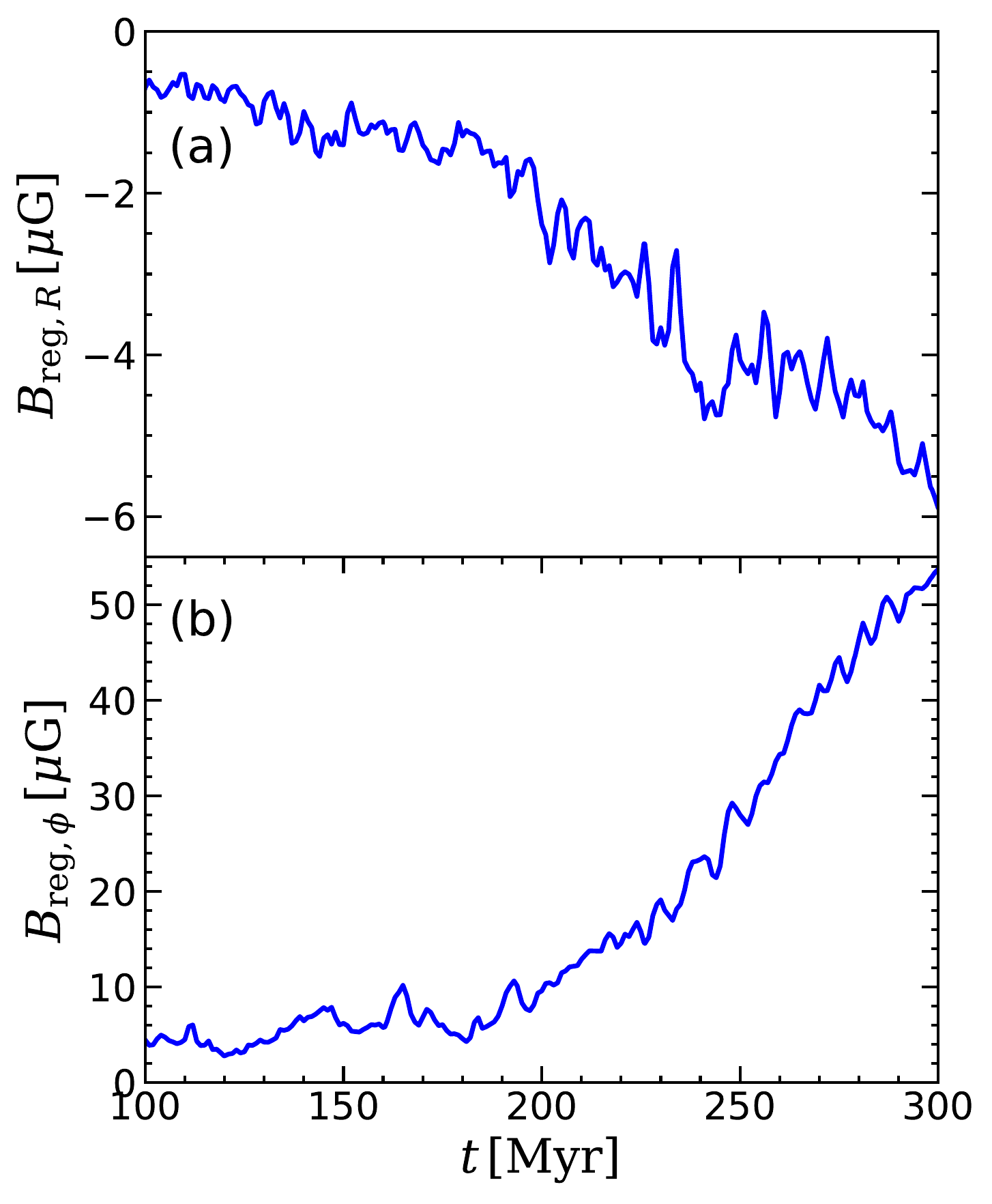}
    \caption{Temporal changes of the (a) radial and (b) azimuthal components of $\mathbf{B}_\mathrm{reg}$ for model \texttt{B100}. Note that $B_{\mathbf{reg}, R}<0$, i.e., it points toward the center. The pitch angle $\theta_p \equiv - \tan^{-1}(B_{\mathrm{reg},R}/B_{\mathrm{reg},\phi})$ is roughly constant at $\theta_p\sim 12^\circ$ for $t=100$--$240\,\mathrm{Myr}$ and $\sim 6^\circ$ for $t=270$--$300\,\mathrm{Myr}$.}
    \label{fig:bfield_history_B100}
\end{figure}

The magnetic fields inside the ring and its interior  are close to axisymmetric (\cref{fig:evolution}), which motivates us to decompose the fields into a regular component $\overline{\mathbf{B}}$ and a irregular, turbulent component $\delta\mathbf{B}$ as
\begin{equation}\label{eq:bfield-decomposition}
    \mathbf{B}(R,\phi,z) = \overline{\mathbf{B}}(R,z) + \delta\mathbf{B}(R,\phi,z),
\end{equation}
where the overbar denotes an azimuthal average
\begin{equation}\label{eq:azimuthal_average}
    \overline{X}(R,z) \equiv \frac{1}{2\pi}\int_0^{2\pi} X\,d\phi,
\end{equation}
for any physical quantity $X$.
Note that $\overline{\delta\mathbf{B}}=0$ by definition.

\Cref{fig:rzplane} plots the spatial distributions of the azimuthally-averaged hydrogen number density $\overline{n}_\mathrm{H}$, the strength of the regular component  $|\overline{\mathbf{B}}| \equiv (\overline{B}_R^2 + \overline{B}_\phi^2 + \overline{B}_z^2)^{1/2}$, and the strength of the turbulent component $|\overline{\delta \mathbf{B}^2}|^{1/2} \equiv (\overline{\delta B_R^2} + \overline{\delta B_\phi^2} + \overline{\delta B_z^2})^{1/2}$ in the $R$--$z$ plane, for model \texttt{B100} at $t=200$ and $300\,\mathrm{Myr}$.
At $t=200\,\mathrm{Myr}$, gas and magnetic fields are concentrated mostly in the nuclear ring delineated by the black circles centered at $(R,z) = (500,0)\,\mathrm{pc}$ with radius $200\,\mathrm{pc}$, while the region outside the ring is filled with diffuse gas.
At this time, magnetic fields are dominated by the turbulent component, especially outside the rings: the density-weighted (see below) mean strength of the regular and turbulent components are $8$ and $22\,\mu\mathrm{G}$, respectively.
The \ac{CND} that begins to form near the center at $t\sim 250\,\mathrm{Myr}$ due to magnetically-driven accretion from the ring is visible in the $t=300\,\mathrm{Myr}$ panels in the right column of \cref{fig:rzplane}, as marked by the rectangles.
The \ac{CND} in our models is denser and more strongly magnetized than the ring.
At $t=300\,\mathrm{Myr}$, the bottom two panels show that the regular magnetic field is stronger than the turbulent field in the ring.

To quantify the magnetic fields within the ring, we define the density-weighted average of the regular, turbulent, and total magnetic fields as
\begin{equation}\label{eq:breg}
    B_{\mathrm{reg},j}\equiv \frac{\iint \overline{\rho} \overline{B_j}\,dRdz}{\iint \overline{\rho}\,dRdz},
\end{equation}
\begin{equation}
    B_{\mathrm{trb},j}\equiv \frac{\iint \overline{\rho} \overline{\delta B_j^2}^{1/2}\,dRdz}{\iint \overline{\rho}\,dRdz},
\end{equation}
\begin{equation}
    B_{\mathrm{tot},j}\equiv \frac{\iint \overline{\rho} \overline{B_j^2}^{1/2}\,dRdz}{\iint \overline{\rho}\,dRdz},
\end{equation}
where the integration is performed over the circular regions shown in  \cref{fig:rzplane}.
Note that $\overline{B_j^2} = \overline{B_j}^2+\overline{\delta B_j^2}$ by definition.
\Cref{fig:bfield_history} plots the time evolution of ${B}_\mathrm{reg}=|\mathbf{B}_\mathrm{reg}|$, ${B}_\mathrm{trb}=|\mathbf{B}_\mathrm{trb}|$, and ${B}_\mathrm{tot}=|\mathbf{B}_\mathrm{tot}|$ for all models.
We note that the ring is quite eccentric and undergoes damped oscillations of eccentricity before it enters a quasi-steady state at $t\sim 100\,\mathrm{Myr}$,
in which case  $\overline{\mathbf{B}}$ and $\delta\mathbf{B}$ do not properly represent the regular and turbulent components.\footnote{
During its eccentricity oscillations, the ring become almost circular at $t\sim 60\,\mathrm{Myr}$ temporarily, producing the peak of $B_\mathrm{reg}$, $B_\mathrm{trb}$, and $B_\mathrm{tot}$ at that time.}

The regular fields grow superlinearly in time (neglecting temporal fluctuations), reaching $B_\mathrm{reg}\sim 50$--$70\,\mu\mathrm{G}$ at the end of the simulations.
In contrast, the turbulent fields  grow initially but saturate at $B_\mathrm{trb} \sim 30$--$40\,\mu\mathrm{G}$.
The total magnetic fields are initially dominated by the turbulent component, but are overtaken by the regular component at later time.
The growth rate of the regular magnetic field at late time is largely insensitive to  $\beta_\mathrm{in}$, suggesting again that the field amplification is not due to the advection of magnetic energy through the nozzles (see  \autoref{app:magnetic-energy-conservation}).
The magnetic fields grow earlier in models with smaller $\beta_\mathrm{in}$ because of the stronger seed fields.
The growth of magnetic fields is most likely driven by \ac{SN} feedback and rotational shear, which we will discuss in \autoref{s:dynamo}.

\Cref{fig:bfield_history_B100} plots the temporal changes of the radial and azimuthal components of $\mathbf{B}_\mathrm{reg}$ for model \texttt{B100} (the vertical component of $\mathbf{B}_\mathrm{reg}$ is negligible), showing that both components grow in time.
The sign of the radial component is the opposite of the sign of the azimuthal field, implying a trailing spiral geometry (see \cref{fig:evolution}), consistent with observed large-scale magnetic fields in the nuclear ring of NGC 1097 \citep{beck99, beck05, lopez-rodriguez21}.
However, the pitch angle of the regular fields $\theta_p \equiv - \tan^{-1}(B_{\mathrm{reg},R}/B_{\mathrm{reg},\phi})$ is $\sim 6^\circ$--$12^\circ$, much smaller than $\theta_p \sim 40^\circ$ inferred from the observations; indeed, we would not expect that the field geometry probed by synchrotron emission would be directly comparable to the mass-weighted magnetic field we show here.

\subsection{Effects of Magnetic Fields on Star Formation}\label{s:effects-of-bfields-on-sf}

\begin{figure}
    \plotone{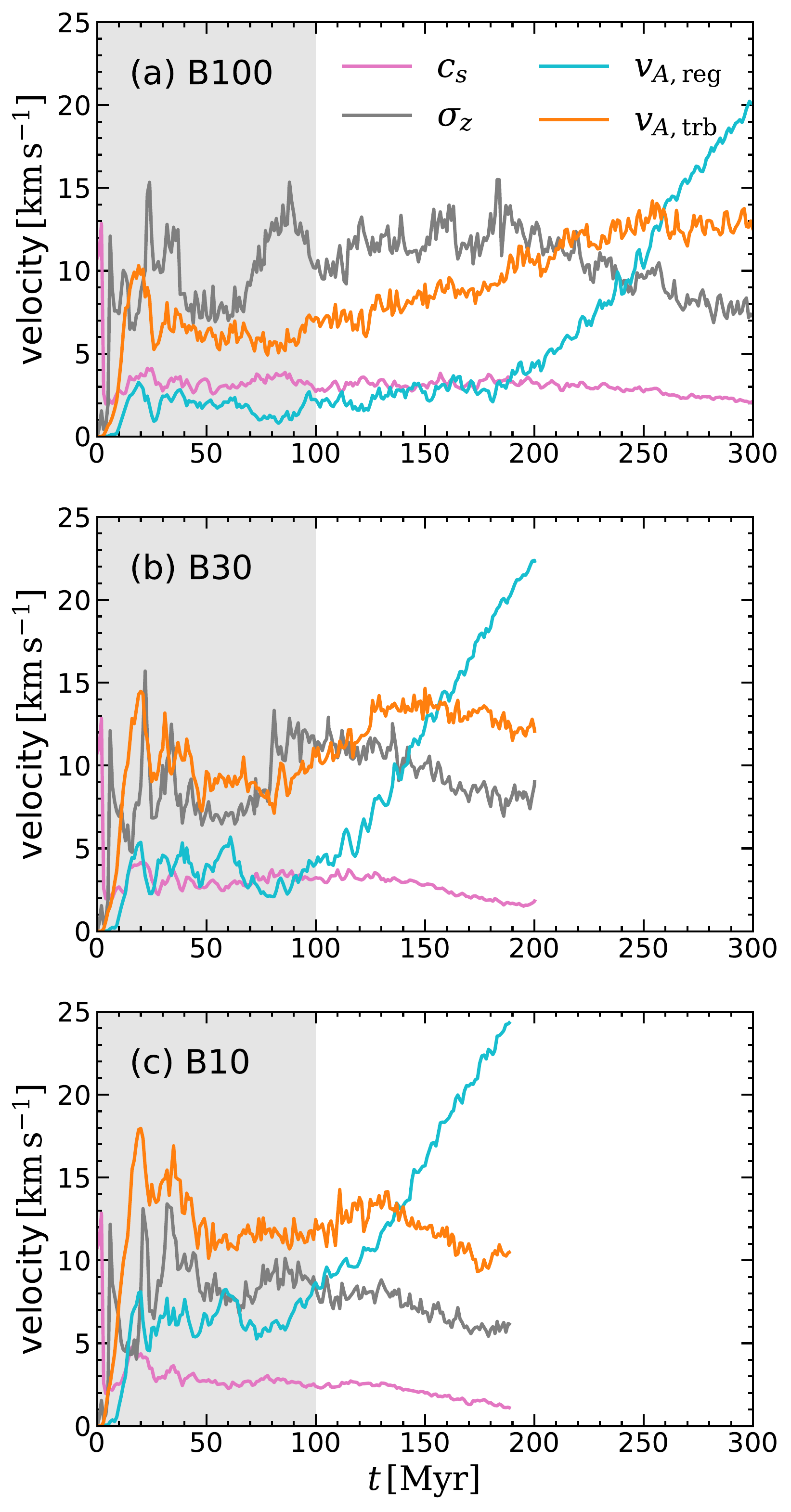}
    \caption{Temporal histories of the sound speed $c_s$ (pink), the vertical velocity dispersion $\sigma_z$ (gray), the Alfv\'en speed associated with the regular $v_{A,\mathrm{reg}}$ (cyan) and turbulent $v_{A,\mathrm{trb}}$ (orange) magnetic fields for models (a) \texttt{B100}, (b) \texttt{B30}, and (c) \texttt{B10}.
    The shaded region represents the epoch when the ring is not fully circularized.}
    \label{fig:velocities}
\end{figure}

\begin{figure*}
    \plotone{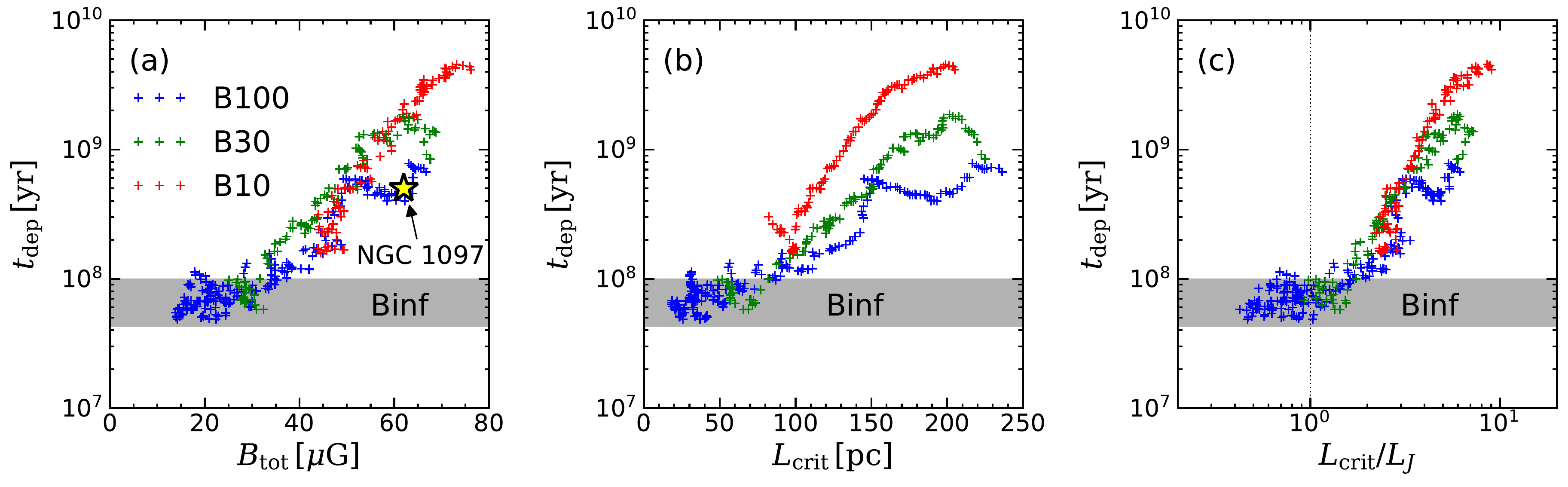}
    \caption{
    Gas depletion time as a function of (a) the total magnetic field strength in the ring, (b) the magnetic critical length $L_\mathrm{crit} = G^{-1/2}R_\mathrm{ring}(\Phi_M/M)$ (see text), and (c) a dimensionless ratio $L_\mathrm{crit}/L_J$, where $L_J=[\pi c_s^2/(G\rho)]^{1/2}$ is the Jeans length of the average cold--warm medium in the ring. Blue, green, and red symbols correspond to models \texttt{B100}, \texttt{B30}, and \texttt{B10}, respectively.
    The yellow star in the panel (a) marks the observed values for the nuclear ring of NGC 1097 \citep{tabatabaei18,prieto19}.
    The vertical dotted line in the panel (c) marks $L_\mathrm{crit} = L_J$.
    The shaded region represents the range of the depletion time in model \texttt{Binf} for $t=100$--$200\,\mathrm{Myr}$.}
    \label{fig:quenching}
\end{figure*}

It is well known that magnetic fields inhibit star formation by providing additional pressure and tension to resist gravitational collapse \citep[e.g.][]{mestel56, mckee95, hennebelle19, jgkim21}.
To assess the dynamical importance of magnetic fields relative to thermal and turbulent pressures, we measure the sound speed $c_s$, vertical velocity dispersion $\sigma_z$, Alfv\'en speed associated with regular $v_{A,\mathrm{reg}}$ and turbulent $v_{A,\mathrm{trb}}$ magnetic fields of the cold--warm medium with $T<2\times 10^4\,\mathrm{K}$ at the midplane as
\begin{equation}\label{eq:cs}
    c_s = \left(\frac{\iiint_{z=-\Delta z}^{z=\Delta z} P\Theta\,dxdydz}{\iiint_{z=-\Delta z}^{z=\Delta z}\rho\Theta\,dxdydz}\right)^{1/2},
\end{equation}
\begin{equation}\label{eq:sz}
    \sigma_z = \left(\frac{\iiint_{z=-\Delta z}^{z=\Delta z} \rho v_z^2\Theta\,dxdydz}{\iiint_{z=-\Delta z}^{z=\Delta z}\rho\Theta\,dxdydz}\right)^{1/2},
\end{equation}
\begin{equation}\label{eq:vA_reg}
    v_{A,\mathrm{reg}} = \left(\frac{\iiint_{z=-\Delta z}^{z=\Delta z} |\overline{\mathbf{B}}|^2\Theta\,dxdydz}{4\pi\iiint_{z=-\Delta z}^{z=\Delta z}\rho\Theta\,dxdydz}\right)^{1/2},
\end{equation}
\begin{equation}\label{eq:vA_trb}
    v_{A,\mathrm{trb}} = \left(\frac{\iiint_{z=-\Delta z}^{z=\Delta z} |\overline{\delta\mathbf{B}^2}|\Theta\,dxdydz}{4\pi\iiint_{z=-\Delta z}^{z=\Delta z} \rho\Theta\,dxdydz}\right)^{1/2},
\end{equation}
where $\Theta = 1$ for $T < 2\times 10^4\,\mathrm{K}$ and $0$ otherwise (see \cref{eq:bfield-decomposition} and related text for definitions of $|\overline{\mathbf{B}}|$ and $|\overline{\delta\mathbf{B}^2}|$).
Note that the integration in the horizontal directions is performed over the annular regions between $R_\mathrm{min}=300\,\mathrm{pc}$ and $R_\mathrm{max}=700\,\mathrm{pc}$, i.e., the ring regions defined in  \cref{fig:rzplane}.

\Cref{fig:velocities} plots temporal histories of $c_s$, $\sigma_z$, $v_{A,\mathrm{reg}}$, and $v_{A,\mathrm{trb}}$ from the magnetized models, showing that $c_s\sim 3\,\mathrm{km\,s^{-1}}$ and $\sigma_z \sim 10\,\mathrm{km\,s^{-1}}$ with modest variations with time.
The turbulent Alfv\'en speed saturates at a level where the turbulent magnetic energy is roughly comparable to the kinetic energy, with $v_{A,\mathrm{trb}} \sim (1.5$--$1.8) \sigma_z$.
The ratio of turbulent magnetic to turbulent kinetic energy found in previous simulations \citep{cgkim15,cgkim17,pakmor17,ostriker22} is in the range ~0.2--0.5, similar or perhaps slightly below what we find here.
In contrast, the regular Alfv\'en speed is initially constant, but exhibits secular growth toward the end of each run, eventually overtaking $v_{A,\mathrm{trb}}$ and reaching $v_{A,\mathrm{reg}} > 20\,\mathrm{km\,s^{-1}}$.
The value of $v_{A,\mathrm{reg}}$ begins to rise at $\sim 200 {\mathrm{Myr}}$ in model {\tt{B100}}, and at $\sim 120 {\mathrm{Myr}}$ in models {\tt{B30}} and {\tt{B10}}.
Intriguingly, the period of rising $v_{A,\mathrm{reg}}$ concides with the period of rising $t_\mathrm{dep}$ for each model (see \cref{fig:history}).

\Cref{fig:quenching}(a) plots the gas depletion time as a function of $B_\mathrm{tot}$, showing that $t_\mathrm{dep}$ has a positive correlation with $B_\mathrm{tot}\gtrsim 30\,\mu\mathrm{G}$, while it is almost independent of $B_\mathrm{tot}\lesssim 30\,\mu\mathrm{G}$.
When there are strong toroidal magnetic fields, a portion of a ring cannot collapse perpendicular to the magnetic fields unless it becomes massive enough by gathering mass along the field line.
The minimum length that has to collapse along the field to become magnetically supercritical is given by
\begin{equation}
 L_\mathrm{crit} \equiv \frac{B}{2\pi  G^{1/2}\rho} = \frac{R_\mathrm{ring}}{G^{1/2}} \left(\frac{\Phi_M}{M}\right),
\end{equation}
where $\Phi_M/M$ is the flux-to-mass ratio of the ring \citep{mestel56,chen14}.
We calculate $\Phi_M/M$ inside the ring region defined in \cref{fig:rzplane} using $\Phi_M = \iint \overline{B}_\phi\,dRdz$.
\Cref{fig:quenching}(b) shows that $t_\mathrm{dep}$ has an overall positive correlation with $L_\mathrm{crit}$, although different models have different $t_\mathrm{dep}$ at a given $L_\mathrm{crit}$.
\Cref{fig:quenching}(c) plots $t_\mathrm{dep}$ against a dimensionless ratio $L_\mathrm{crit}/L_J$, where $L_J \equiv [\pi c_s^2/(G\rho)]^{1/2}$ is the Jeans length of the cold--warm medium in the ring\footnote{For typical \emph{ambient} cold--warm medium density, we take the volume-averaged density of the cold--warm medium between $R=300$ and $700\,\mathrm{pc}$ at the midplane, with the density cut $n_\mathrm{H}>10\,\mathrm{cm}^{-3}$ to exclude the warm, tenuous gas in the inflowing streams.
For model \texttt{B100} before $t\sim 200\,\mathrm{Myr}$, this is $\sim 100\,\mathrm{cm}^{-3}$, a factor of $\sim 4$ smaller than the mass-weighted mean.}, showing that $t_\mathrm{dep}$ is approximately constant for $L_\mathrm{crit}/L_J \lesssim 1$ but increases with $L_\mathrm{crit}/L_J \gtrsim 1$.
This is because as $L_\mathrm{crit}$ exceeds $L_J$, more and more Jeans-unstable clumps (smallest ones first) fail to ultimately collapse because they are magnetically subcritical.
We note that at later time the \ac{MJI} operates to gather material along the ring circumference.
However, when $L_\mathrm{crit}$ exceeds the typical spacing $\sim 100\,\mathrm{pc}$ between the spiral segments formed by the \ac{MJI}, even the mass gathered by the \ac{MJI} is not enough to overcome magnetic support.
Overall, \cref{fig:quenching} suggests that large-scale magnetic fields tend to suppress star formation in nuclear rings.

\subsection{Vertical Dynamical Equilibrium}

\begin{figure}
    \plotone{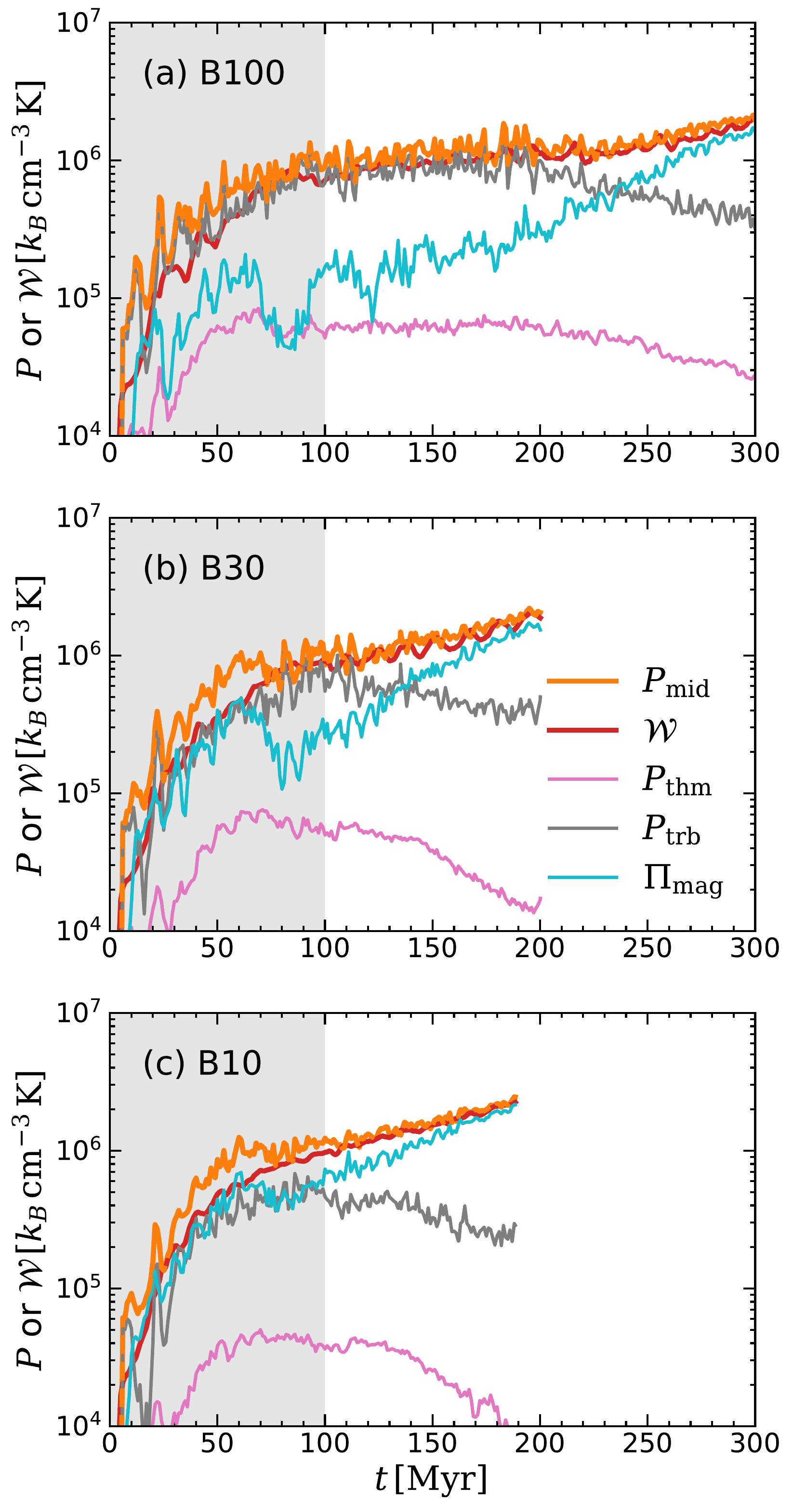}
    \caption{
    Temporal histories of the total midplane pressure (orange) and gas weight (red), as well as the thermal (pink), turbulent (gray), and magnetic (cyan) components of the midplane pressure, for models (a) \texttt{B100}, (b) \texttt{B30}, and (c) \texttt{B10}. After the ring enters the steady state at $t \sim 100\,\mathrm{Myr}$, $P_\mathrm{mid}\approx \mathcal{W}$, indicating that the vertical dynamical equilibrium holds very well.
    The shaded region represents the epoch when the ring is not fully circularized.}
    \label{fig:vertical-equilibrium}
\end{figure}

A framework developed to understand the co-regulation of
galactic \acp{SFR} and ISM properties is the \ac{PRFM} theory (\citealt{ostriker22}; see also \citealt{ostriker10}; \citealt{ostriker11}).
The theory assumes that the \ac{ISM} in disk galaxies satisfies vertical dynamical equilibrium between the total midplane pressure and the weight of the overlying gas, and that it is star formation feedback that heats the gas and drives turbulence to maintain the required level of the midplane pressure.
In this picture, the \ac{SFR} is determined by the requirement for the feedback to yield the pressure needed for vertical dynamical equilibrium, which not only depends on the gas surface density but also on the local stellar density and the velocity dispersion (or gas scale height).
Here we check if the vertical dynamical equilibrium holds in the magnetized nuclear ring, and assess the relative importance of the magnetic pressure to the other pressures.

We measure the thermal, turbulent, and magnetic pressures of the cold--warm medium at the midplane as
\begin{equation}
    P_\mathrm{thm} = \frac{\iiint_{z=-\Delta z}^{z=\Delta z} P\Theta\,dzdxdy}{\iiint_{z=-\Delta z}^{z=\Delta z} \Theta\,dzdxdy},
\end{equation}
\begin{equation}
    P_\mathrm{trb} = \frac{\iiint_{z=-\Delta z}^{z=\Delta z} \rho v_z^2\Theta\,dzdxdy}{\iiint_{z=-\Delta z}^{z=\Delta z} \Theta\,dzdxdy},
\end{equation}
\begin{equation}
    \Pi_\mathrm{mag} = \frac{\iiint_{z=-\Delta z}^{z=\Delta z} T_{zz}\Theta \,dzdxdy}{\iiint_{z=-\Delta z}^{z=\Delta z} \Theta \,dzdxdy},
\end{equation}
where the integration in the horizontal directions is performed over the ring regions between $R_\mathrm{min}=300\,\mathrm{pc}$ and $R_\mathrm{max}=700\,\mathrm{pc}$, and $\Theta = 1$ for $T < 2\times 10^4\,\mathrm{K}$ and $0$ otherwise.
Note that $T_{zz} = B^2/(8\pi) - B_z^2/(4\pi)$ so that $\Pi_\mathrm{mag}$ represents the total vertical magnetic stress, including both magnetic pressure and tension terms \citep{boulares90,piontek2007,cgkim15}.
The weight of the \ac{ISM} is given by
\begin{equation}
    \mathcal{W} = \frac{1}{A_\mathrm{ring}}\iiint_{z=0}^{z=L/2}\rho\frac{\partial\Phi_\mathrm{tot}}{\partial z}\,dz dxdy,
\end{equation}
where the horizontal integration is performed over the ring region as before, and $A_\mathrm{ring}\equiv \pi(R_\mathrm{max}^2 - R_\mathrm{min}^2)$.
It follows from \cref{eq:momentum} that under quasi-steady equilibrium, $P_\mathrm{mid}\equiv P_\mathrm{thm}+P_\mathrm{trb}+\Pi_\mathrm{mag}\approx\mathcal{W}$ if the pressures at the horizontal and the upper boundaries of the cylindrical annulus are small compared to the midplane value.

\Cref{fig:vertical-equilibrium} plots $P_\mathrm{mid}$ and $\mathcal{W}$ as well as the contributions of each pressure component for magnetized models, showing $P_\mathrm{mid}\approx\mathcal{W}$ indeed holds well once the ring enters the quasi-steady state at $t\sim 100\,\mathrm{Myr}$ \footnote{The average midplane pressure including hot ($T>2\times 10^4\,\mathrm{K}$) gas very well matches the weight for all time, even before $t\sim 100\,\mathrm{Myr}$; For $t<100\,\mathrm{Myr}$, the average midplane pressure of the cold--warm medium is somewhat higher than the weight, indicating the hot gas pressure at those times is slightly smaller than that of the cold--warm medium.}.
While the midplane pressure is dominated by the turbulent component at early time, the magnetic pressure dominates after $t\sim 240\,\mathrm{Myr}$, $\sim 140\,\mathrm{Myr}$, and $\sim 100\,\mathrm{Myr}$ for models \texttt{B100}, \texttt{B30}, and \texttt{B10}, respectively.
As the ring becomes magnetically supported against the vertical gravity, the demand for the stellar feedback to replenish the thermal and turbulent pressures diminishes, causing the \ac{SFR} to decline (\cref{fig:history}a),  consistent with the \ac{PRFM} theory.

\subsection{Interpretation of the Field Growth}\label{s:dynamo}

In our simulations, both regular and turbulent fields grow in strength with time, although the latter saturates at $\sim 35\,\mu\mathrm{G}$.
The rapid growth and saturation of the turbulent magnetic fields are likely due to the \ac{SN}-driven turbulence, which not only randomly stretches, twists, and folds the field lines to amplify them at small scales \citep{vainshtein72,childress95}, but also tangles the large-scale field lines to create fluctuating components.
A number of simulations of the \ac{ISM} where the turbulence is naturally driven by the \ac{SN} feedback have demonstrated that the turbulent magnetic fields can be amplified out of very weak seed fields \citep[e.g.,][]{cgkim15,rieder16,rieder17,butsky17,pakmor17,gent21}.
These studies have found that the growth rate of the turbulent dynamo is sensitive to the numerical resolution because the fastest growth occurs at the smallest resolvable scale, although the saturation amplitude is almost independent of the numerical resolution.

Compared to the small-scale dynamo, the large-scale dynamo responsible for the growth of ordered magnetic fields is still poorly understood.
In part, the growth of $B_\mathrm{reg}$ in our simulations is presumably due to the strong differential rotation in the ring, which stretches radial fields into the azimuthal direction.
However, the most naive version of the stretching effect is not evident in our simulations.
For pure differential rotation $\mathbf{v} = \mathbf{v}_\mathrm{rot} = R\Omega(R)\mathbf{e}_\phi$, it can be shown from \cref{eq:induction,eq:azimuthal_average}, and $\boldsymbol{\nabla}\cdot\mathbf{B} = 0$ that
\begin{subequations}\label{eq:mean-field0}
\begin{align}
    \frac{\partial \overline{B}_R}{\partial t} &= 0,\label{eq:dBRdt}\\
    \frac{\partial \overline{B}_\phi}{\partial t} &= -q\Omega \overline{B}_R,\label{eq:omega-effect}
\end{align}
\end{subequations}
are exactly satisfied.
Here, $q\equiv -d\ln\Omega/d\ln R$ is the rate of shear.
In our simulations, $q = 0.87$ and $\Omega = 0.45\,\mathrm{Myr}^{-1}$ at $R = R_\mathrm{ring}$.
Taking $\overline{B}_R = -1\,\mu\mathrm{G}$ as is true for $B_{\mathrm{reg},R}$ at $t\sim 100{\rm Myr}$ in model {\tt B100}, \cref{eq:omega-effect} would imply a growth of
$\overline{B}_\phi$ from zero to $40\,\mu\mathrm{G}$ in $100\,\mathrm{Myr}$, vastly overestimating the true growth of $B_{\mathrm{reg},\phi}$ shown in \cref{fig:bfield_history_B100}.
In addition, while \cref{eq:dBRdt} would predict $B_{\mathrm{reg},R}$ to be constant in time, \cref{fig:bfield_history_B100} shows the magnitude of $B_{\mathrm{reg},R}$ in fact grows in time.

The discrepancies with respect to the prediction of the simple shear model (\cref{eq:mean-field0})  indicate that velocity components other than $\mathbf{v}_\mathrm{rot}$ play an important role in governing the growth of the regular magnetic fields in our simulations.
Although a quantitative analysis of the large-scale dynamo that is responsible for growth of $\mathbf{B}_\mathrm{reg}$ is beyond the scope of this work, we provide a qualitative account of the mean field growth to motivate future work.
One may write a general velocity field by
\begin{equation}\label{eq:general_vfield}
   \mathbf{v} = \mathbf{v}_\mathrm{rot} + \mathbf{v}_\mathrm{blk} + \delta \mathbf{v},
\end{equation}
where $\mathbf{v}_\mathrm{blk} \equiv \overline{\mathbf{v}} - \mathbf{v}_\mathrm{rot}$  roughly corresponds to bulk motions of the fluid which deviate from circular rotation (e.g., bubble expansion and radial accretion), and $\delta\mathbf{v}$ is random turbulent motion.
Here, $\overline{\mathbf{v}}$ and $\delta \mathbf{v}$ are similarly defined as \cref{eq:bfield-decomposition} such that $\overline{\delta \mathbf{v}} = 0$ by definition.
Substituting \cref{eq:general_vfield} into \cref{eq:induction} and applying the averaging of \cref{eq:azimuthal_average}, one obtains
\begin{equation}\label{eq:mean-field}
    \frac{\partial \overline{\mathbf{B}}}{\partial t} = -q\Omega\overline{B}_R\mathbf{e}_\phi + \boldsymbol{\nabla}\times(\mathbf{v}_\mathrm{blk}\times\overline{\mathbf{B}}) + \boldsymbol{\nabla}\times (\overline{\delta\mathbf{v}\times\delta\mathbf{B}}).
\end{equation}
In the mean-field dynamo theory for rotating systems, it is thought that the last term in \cref{eq:mean-field} captures the so called $\alpha$ effect in which the Coriolis force yields systematic twists in the field lines to produce large-scale magnetic fields, as envisaged by \citet{parker55}, as well as turbulent diffusion of the mean magnetic fields \citep{brandenburg05}.
For example, by assuming weak Lorentz force and isotropic turbulence, the first-order smoothing approximation yields $\boldsymbol{\nabla}\times (\overline{\delta \mathbf{v} \times \mathbf{\delta B}}) = \alpha\boldsymbol{\nabla}\times \overline{\mathbf{B}} + \eta_t \boldsymbol{\nabla}^2\overline{\mathbf{B}}$, with the transport coefficients $\alpha \approx -(\tau_\mathrm{cor}/3)\overline{\mathbf{\delta v}\cdot(\boldsymbol{\nabla}\times\mathbf{\delta v})}$ and $\eta_t \approx (\tau_\mathrm{cor}/3) \overline{\mathbf{\delta v}^2}$, where $\tau_\mathrm{cor}$ is the correlation time of turbulence \citep{brandenburg05}.
If $B_{\mathrm{reg},\phi}$ has a single sign and decreases in magnitude away from the midplane, as is true in our simulations, the radial component of $\boldsymbol{\nabla} \times \overline{\mathbf{B}}$ will change sign across the midplane.
Expansion of bubbles centered on the midplane in combination with the Coriolis force will tend to produce a kinetic helicity $\overline{\mathbf{\delta v}\cdot(\boldsymbol{\nabla}\times\mathbf{\delta v})}$ that changes sign across the midplane \citep{ruzmaikin88}, and indeed we find this sign change when we measure kinetic helicity in our simulations.
Thus, we can qualitatively understand the growth of $B_{\mathrm{reg}, R}$ as due to the combined effect of the midplane change in sign of $\alpha$ with the midplane change in sign of $(\boldsymbol{\nabla} \times \overline{\mathbf{B}})_R$, to make the last term in \cref{eq:mean-field} keep the same sign across the midplane.

As discussed above, the presence of nonzero $\overline{B}_R$ will tend to produce growth of $\overline{B}_\phi$, as expressed by the first term proportional to $\Omega$ on the right-hand side of \cref{eq:mean-field}.
The combination of the two effects discussed above on mean field growth is often referred to as an ``$\alpha$-$\Omega$'' dynamo.
The turbulent diffusion term $\eta_t \boldsymbol{\nabla}^2 \overline{\mathbf{B}}$ will, however, tend to suppress growth of the mean field.
Also, the second term on the right hand side of \cref{eq:mean-field} could in the case of our simulations represent advection of mean magnetic fields out of the ring due to the radial accretion flow, and/or dilution due to the expansion of \ac{SN} remnants.
Together with the turbulent diffusion, these bulk flows may account for the reduction in field growth compared to a pure $\alpha$--$\Omega$ dynamo.
We note that the regular magnetic fields grow faster when $\sigma_z$ starts to decline (see \cref{fig:bfield_history_B100,fig:velocities}(a)), which is presumably due to the reduction of $\eta_t$ with decreasing velocity dispersion \citep{gressel08}.

One important aspect of the $\alpha$ effect is that while it creates large-scale magnetic helicity associated with the poloidal loops, it does so at the expense of a small-scale magnetic helicity of an opposite sign, which is associated with the internal twist of the poloidal loops \citep[see][for a visual illustration]{blackman03}, in a way that satisfies magnetic helicity conservation.
As small scale magnetic helicity (or “twists”) accumulates over time, the magnetic tension resists bending and twisting of the field lines to quench the $\alpha$ effect, potentially limiting the growth of regular magnetic fields \citep[see Section 9 of][]{brandenburg05}.
\citet{shukurov06} showed that galactic fountain flows can transport the small-scale helicity out of the disk in vertical direction, maintaining the dynamo against the back-reaction from the Lorentz force.
In our simulations, clustered \ac{SN} explosions are powerful enough to drive outflows that drag magnetic fields away from the midplane and to leave the computational domain (see \cref{fig:superbubble}; see also Figure 8 of \citetalias{moon21}).
Even when vertical outflows become weak as the growing magnetic fields suppress the \ac{SFR}, the radial accretion flows may still be able to transport helicity out of the ring.

\subsection{Resolution Dependence}\label{s:resolution}

\begin{figure}
    \plotone{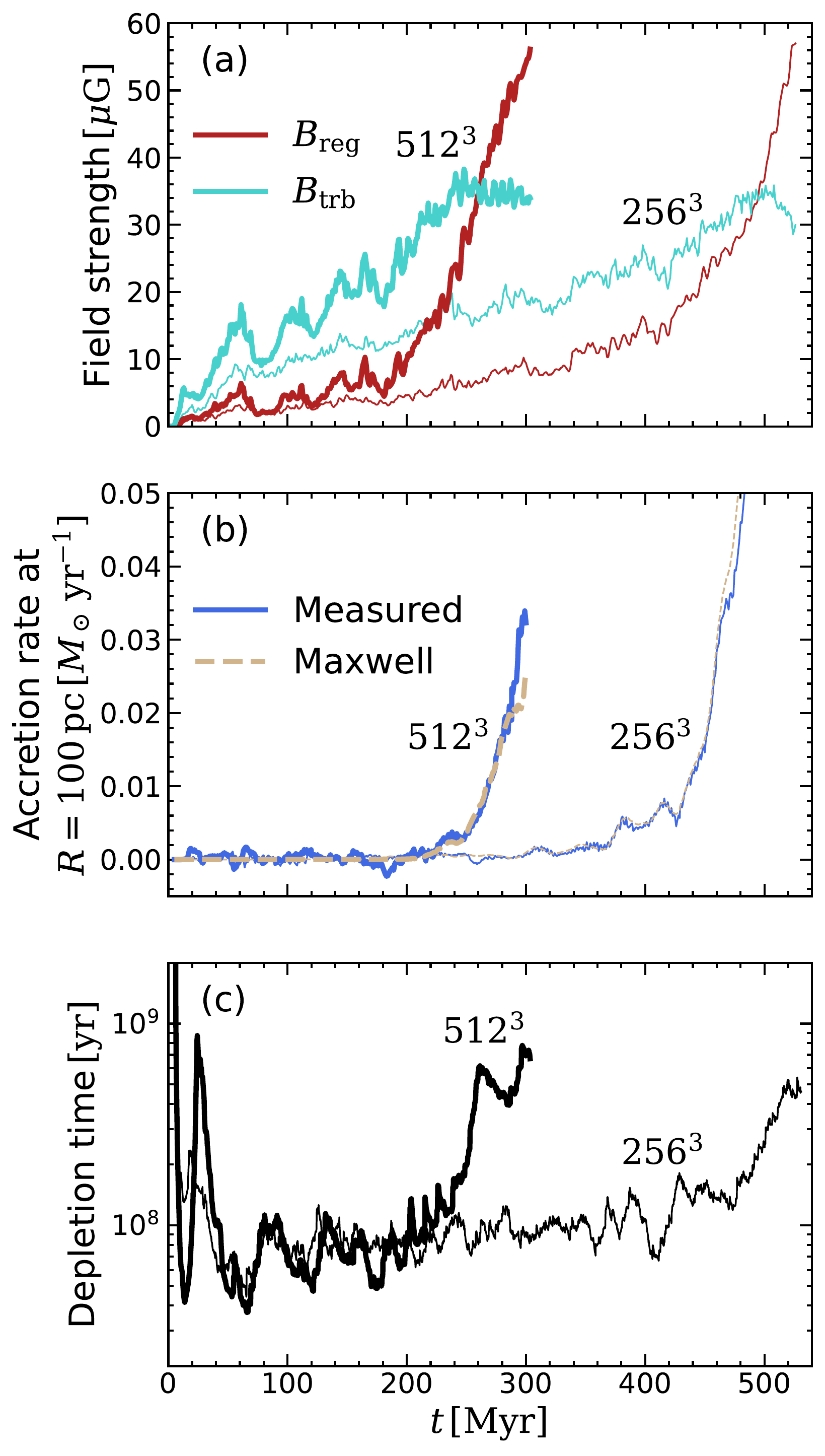}
    \caption{Resolution study. We compare 
    the $512^3$ ($\Delta x = 4\,\mathrm{pc}$; thick) and $256^3$ ($\Delta x = 8\,\mathrm{pc}$; thin) runs with $\beta_\text{in}=100$ for evolution of (a) the regular (brown) and turbulent (cyan) magnetic field strengths, (b) the measured ($\dot{M}_\mathrm{acc}$; blue solid) and predicted ($\dot{M}_\mathrm{M}$; gold dashed) mass accretion rates at $R=100\,\mathrm{pc}$, and (c) the depletion time $t_\mathrm{dep}$.}
    \label{fig:resolution}
\end{figure}

To see how our results depend on numerical resolution, we rerun our fiducial model \texttt{B100} at lower resolution, using $256^3$ cells corresponding to $\Delta x = 8\,\mathrm{pc}$.
\Cref{fig:resolution}(a) compares the temporal evolution  of the field strength between the $512^3$ and $256^3$ runs.
Evidently, growth of the turbulent field $B_{\rm trb}$ is initially higher for the higher-resolution model, and the super-linear growth stage for the regular field $B_{\rm reg}$ also occurs earlier in time for the higher-resolution model.
The initially faster growth of $B_{\rm trb}$ is presumably because the growth rate of the small-scale turbulent dynamo is inversely proportional to the eddy turnover time at the grid scale, as noted by \citet{rieder16}.
The stronger turbulent fields at earlier time in the $512^3$ run also presumably lead to earlier superlinear growth of $B_{\rm reg}$.
We note that the growth rate of $B_{\rm reg}$ at the time when $B_{\rm trb}$ saturates is similar in both models.
The saturated field strength of $B_{\rm trb}$ is similar in both models (see also \autoref{fig:alfven_ceiling} and related discussion of saturation).

Because magnetic field growth is delayed in the low-resolution model, other characteristic evolutionary effects affected by magnetic fields (see \autoref{s:overall-evolution}) also occur later in time.
For example, \cref{fig:resolution}(b) and (c) show that the mass accretion rate and depletion time start to increase at $t\sim 200\,\mathrm{Myr}$ and $\sim 400\,\mathrm{Myr}$ in the $512^3$ and $256^3$ runs, respectively, corresponding to the start of the super-linear growth of $B_{\rm reg}$.
We note that even though the mass accretion rate \emph{at a given time} depends on numerical resolution, it is entirely consistent with the predicted accretion rate from the instantaneous Maxwell stress.

\section{Summary And Discussion}\label{s:summary-discussion}

\subsection{Summary}\label{s:summary}

Nuclear rings at the centers of barred galaxies are active in star formation \citep{mazzuca08,ma18} and threaded by magnetic fields with a mean strength of $\sim 50$--$100\,\mu\mathrm{G}$ \citep{beck05, yang22}.
To study how magnetic fields affect star formation in nuclear rings, we run \ac{MHD} simulations of galactic centers.
We employ the semi-global models of \citetalias{moon21}, in which magnetized gas streams from two nozzles at the boundaries supply gas and magnetic fields, mimicking bar-driven gas inflows along dust lanes.
We adopt the modified TIGRESS framework \citep{cgkim17} to model star formation and related \ac{FUV} and \ac{SN} feedback, as well as the shielding of \ac{FUV} radiation and \ac{CR} heating in dense environments.
We fix the mass inflow rate to $\dot{M}_\mathrm{in}=1\,M_\odot\,\mathrm{yr}^{-1}$ and the ring size to $R_\mathrm{ring}=500\,\mathrm{pc}$, while varying the plasma parameter $\beta_\mathrm{in} = 10, 30, 100$ to adjust the average field strength within the gas streams to $B_\mathrm{in,avg} = 2.4\,\mu\mathrm{G}$, $1.4\,\mu\mathrm{G}$, and $0.76\,\mu\mathrm{G}$.
We also run a hydrodynamic model with unmagnetized streams for comparison.
The magnetic fields in the streams are set parallel to the inflow velocity, motivated by observations \citep{beck05,lopez-rodriguez21}.

The main results of this work can be summarized as follows:
\begin{enumerate}
    \item \emph{Overall Evolution}: The two gas streams injected from the  domain boundaries at the opposite sides collide with each other after about half an orbital time, dissipating their orbital kinetic energy via shocks.
    As the gas orbits gradually circularize, a well-defined nuclear ring forms at the radius where the specific angular momentum of the inflowing gas matches that of the circular orbit.
    At about $t\sim100\,\mathrm{Myr}$, the nuclear ring reaches a quasi-steady state in which the shape, \ac{SFR}, and gas mass become approximately constant with time.
    Stars form randomly across the whole circumference of the ring, and the associated feedback renders the ring turbulent.
    When the magnetic fields in the ring become strong enough they reduce the ring \ac{SFR}.
    At the same time, strong magnetic torques lead to accretion flows from the ring to the galaxy center (\cref{fig:evolution,fig:surf_evolution}), where a circumnuclear disk grows.
    Due to the action of \ac{MJI} combined with strong shear, at late stages the ring forms transient trailing spiral segments, some of which undergo star formation.

    \item \emph{Magnetic Fields and Their Growth}: Magnetic fields in the ring can be separated into a regular and a  turbulent component, where the former is defined via azimuthal averaging and the latter is the azimuthally fluctuating residual.
    The turbulent component grows in strength over time and saturates at $\sim 30$--$40\,\mu\mathrm{G}$ independent of $\beta_\mathrm{in}$, likely due to the \ac{SN} driven turbulent dynamo.
    In contrast, the regular component does not saturate but keeps growing with time, reaching $50$--$70\,\mu\mathrm{G}$ at the end of the runs (\cref{fig:bfield_history}).
    While the turbulent fields are approximately isotropic, the regular fields are dominated by the azimuthal component with a pitch angle of $\theta_p \sim 6^\circ$--$12^\circ$.
    The overall field direction is mostly toroidal near the midplane, but expansion of superbubbles created by clustered \ac{SN}e drag the toroidal fields to produce poloidal fields in high-altitude regions (\cref{fig:superbubble,fig:volume-rendering}).

    \item \emph{Magnetically Driven Accretion}: All our magnetized models develop accretion flows that slowly fill the region interior to the ring and eventually form a \ac{CND} with radius $\lesssim 50\,\mathrm{pc}$ at the center.
    This is in stark contrast to the unmagnetized model where the region interior to the ring is always filled with hot, rarefied gas.
    The gas accretion rates measured in the simulations are consistent with the theoretical quasi-steady rates due to the Maxwell stress, indicating that the radial accretion is driven by magnetic tension.
    The measured accretion rate depends on the galactocentric radius and reaches $\sim (0.02$--$0.1)\,M_\odot\,\mathrm{yr}^{-1}$ at late time (\cref{fig:inflow_rate_t,fig:inflow_rate_R}).

    \item \emph{Effects of Magnetic Fields on Star Formation}: When strong magnetic fields develop in the ring, they suppress star formation therein.
    Consequently, the gas depletion time $t_\mathrm{dep}$ in the rings increases with the total field strength.
    In particular, strong regular azimuthal magnetic fields in the ring limit the radial and vertical compression that lead to collapse, unless the ring undergoes sufficient azimuthal contraction that could gather material along the field lines (\cref{fig:quenching}).
    The ring maintains vertical dynamical equilibrium instantaneously, meaning that the weight of the \ac{ISM} is balanced by the midplane total pressure (\cref{fig:vertical-equilibrium}).
    While the magnetic pressure is negligible in the vertical force balance at early time, it becomes dominant at late times ($t\gtrsim240\,\mathrm{Myr}$ in model \texttt{B100}).
    This late-time strong magnetic support, which is mostly from the regular (non-turbulent) component of the magnetic fields, reduces the demand for \ac{SN} feedback to replenish the thermal and turbulent pressures, thereby indirectly lowers the \ac{SFR}, consistent with the \ac{PRFM} theory of \citet{ostriker22}.
\end{enumerate}

\subsection{Discussion}\label{s:discussion}

Infrared polarization observations indicate that Galactic magnetic fields are preferentially torodial near the \ac{CMZ} and poloidal in the regions with Galactic latitudes $|b|> 0.4^\circ$ \citep{nishiyama10}.
This toroidal--to--poloidal transition of the Galactic magnetic fields is consistent with our numerical results that expanding superbubbles drag the toroidal fields in the rings to produce poloidal magnetic fields in high-altitude regions (\cref{fig:superbubble,fig:volume-rendering}).
The poloidal magnetic walls of venting superbubbles are likely illuminated by relativistic particles accelerated in-situ at \ac{SN} shocks, potentially creating some nonthermal radio filaments, such as Radio Arc and Sgr C filaments, found near radio bubbles \citep[e.g.,][]{heywood22}.
For filaments without any evident source, \citet{barkov19} proposed that transiting pulsar wind nebulae may inject relativistic particles to make the background poloidal magnetic fields visible locally.
Alternatively, \citet{sofue23} proposed that the filaments represent projected wavefronts of fast \ac{MHD} waves launched from \ac{SN}e exploding in the ring, which locally compress the existing magnetic fields.
The results of our simulations suggest that the background volume-filling magnetic fields necessary in both scenarios can be produced by superbubbles breaking out of the ring.

The results of our simulations show that the magnetic torque produces significant inflows of gas from a nuclear ring toward the center, forming a \ac{CND}.
The mass accretion rate depends both on time and radius such that for model \texttt{B100} at $t=300\,\mathrm{Myr}$, $\dot{M}_\mathrm{acc}\sim0.02$--$0.03\,M_\odot\,\mathrm{yr}^{-1}$ near $R\sim 50\,\mathrm{pc}$ while $\sim 0.1\,M_\odot\,\mathrm{yr}^{-1}$ near $R\sim 500\,\mathrm{pc}$.
\Cref{fig:inflow_rate_R} suggests that the radial profile of the mass accretion rate becomes flatter over time, in a way that the accretion rate near the \ac{CND} increases with time.
At $t=300\,\mathrm{Myr}$, the \ac{CND} formed in model \texttt{B100} has gas mass $4\times 10^5\,M_\odot$, which is order of magnitude smaller than the observed \ac{CND} masses $\sim 10^7\,M_\odot$ in nearby active galaxies \citep{combes19}.
We note, however, that the continued mass accretion with the rate of $\sim 0.01$--$0.1\,M_\odot\,\mathrm{yr}^{-1}$ is capable of producing $10^7\,M_\odot$ \acp{CND} within $0.1$--$1\,\mathrm{Gyr}$.
We also note that, because the magnetically driven mass accretion rate is proportional to $-B_R B_\phi = (B_R^2+B_\phi^2)^{1/2} \sin(2\theta_p)/2$, the accretion rate would be higher if the magnetic fields are more loosely wrapped (i.e., larger pitch angle).

\citet{tabatabaei18} measured the magnetic field strength in 11 giant clumps in the nuclear ring of NGC 1097 and found a negative correlation between the star formation efficiency and magnetic field strength, suggesting that magnetic fields are inhibiting star formation in the nuclear ring.
\Cref{fig:quenching}(a) compares
the observed average magnetic field strength of $62\,\mu\mathrm{G}$ \citep{tabatabaei18} and depletion time of $5\times 10^8\,\mathrm{yr}$ \citep{prieto19} with the results of our simulations.
Based on our numerical results, the nuclear ring in NGC 1097 is in the regime where magnetic fields are dynamically important to suppression of star formation.

Our simulations demonstrate the intriguing possibility of star formation quenching in nuclear rings together with mass accretion to the center, both resulting from the growth of large-scale regular magnetic fields.
However, it is still uncertain how the field strength and pitch angles depend on the underlying rotation curve and feedback physics as well as the magnetization of the inflowing gas, which models a bar-driven stream.
These issues may be resolved by running more realistic simulations with improved physics.
First of all,  while the present work assumes that magnetic fields in the gas streams are kept constant, in real galaxies both the field strength and direction will vary with time, resulting in significant changes in the field strength and structure in nuclear rings and inward.
In particular, the magnetic fields in the inflowing stream may change their polarity over a dynamical time at the bar end region and then be amplified in the ring.
Reconnection with existing fields of the opposite polarity in the ring would then prevent excessive growth of the magnetic fields.
To address this issue, it will be necessary to run global simulations of barred galaxies in which gas streams along dust lanes are modeled self-consistently by nonlinear interactions with a bar potential.

\begin{figure}
    \plotone{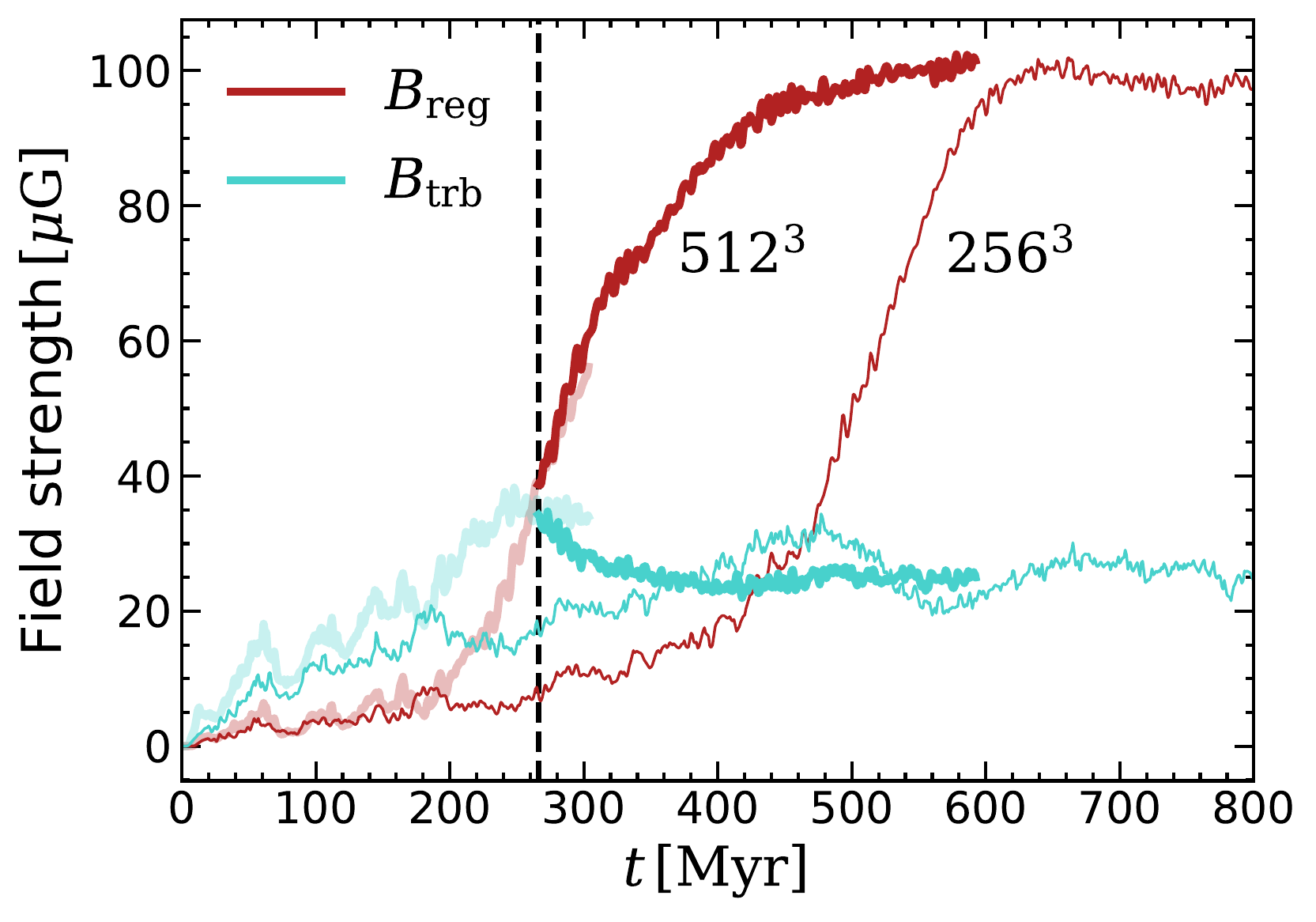}
    \caption{Evolution of the regular (brown) and turbulent (cyan) magnetic field strengths in the $\beta_\mathrm{in}=100$ model with an Alfv\'en speed ceiling applied in order to extend the simulation run time beyond that of model \texttt{B100}. Thick and thin lines correspond to the runs with $512^3$ and $256^3$ cells, respectively. The $512^3$ run is restarted from model \texttt{B100} (light-colored) at $t=266\,\mathrm{Myr}$ marked by the vertical dashed line, while the $256^3$ run has an Alfv\'en ceiling applied throughout its evolution.}
    \label{fig:alfven_ceiling}
\end{figure}

As \Cref{fig:bfield_history} shows, the regular magnetic fields $B_\mathrm{reg}$ in our simulations are still growing at the end of the runs, even though the turbulent component of the magnetic field has saturated.  It is, however, of much interest to know how $B_{\rm reg}$ would evolve over longer timescales.  For our standard simulations, following this late-time evolution is precluded by an extremely short timestep, as limited by the large Alfv\'en speed in locations where the density becomes low.\footnote{Large Alfv\'en speeds in our simulations occur in cones surrounding the $z$-axis above and below the \ac{CND}, which have very low density but moderate magnetic field strength (see the right column of \cref{fig:rzplane}).}
However, by applying an artificial density floor to put a ceiling on the Alfv\'en speed of $10^3\,\mathrm{km\,s^{-1}}$, we are able to restart our fiducial model \texttt{B100} from $t=266\,\mathrm{Myr}$ and run up to $t=600\,\mathrm{Myr}$, as shown in \cref{fig:alfven_ceiling}.  
This figure shows that the magnetic fields saturate at $B_\mathrm{reg} \sim 100\,\mu\mathrm{G}$ around $t\sim 520\,\mathrm{Myr}$.
Additionally, we find that for a corresponding lower-resolution run, the saturation level of $B_{\rm reg}$ is the same, although saturation occurs at a later time (see \autoref{s:resolution}).  
It is difficult to pinpoint what physical effect terminates the growth of $B_\mathrm{reg}$, but radial spreading of the ring, nonlinear quenching due to small-scale helicity, and turbulent diffusion may all play a role \citep[e.g.][]{2022arXiv221103476B}.

Our semi-global framework allows us to afford relatively high resolution of $\Delta x = 4\,\mathrm{pc}$ uniformly across the entire domain such that the expansion of individual \ac{SN} remnants is captured self-consistently.
We note, however, that since the free-fall time of star-forming clumps in the ring is shorter than the \ac{SN} delay time of $\sim 4\,\mathrm{Myr}$, early feedback in the form of radiation and stellar winds may affect gas dynamics and star formation efficiency of individual star-forming clumps significantly in extreme environments like nuclear rings.
As a final caveat, we note that considering the low ionization fraction in densest star-forming clumps in nuclear rings, non-ideal terms in the induction equation that we have neglected in this study might potentially affect the growth of magnetic fields in the ring and therefore star formation and mass accretion flows.

\begin{acknowledgments}
We are grateful to the referee for an insightful report. The work of SM was supported by an NRF (National Research Foundation of Korea) grant funded by the Korean government (NRF-2017H1A2A1043558-Fostering Core Leaders of the Future Basic Science Program/Global Ph.\ D.\ Fellowship Program).
The work of W.-T.K.\ was supported by the grant of National Research Foundation of Korea (2022R1A2C1004810).
The work of C.-G.K.\ was supported in part by NASA ATP grant No.\ 80NSSC22K0717.
The work of ECO is partly supported by the Simons Foundation under grant 510940.
Computational resources for this project were provided by Princeton Research Computing, a consortium including PICSciE and OIT at Princeton University, and by the Supercomputing Center/Korea Institute of Science and Technology Information with supercomputing resources including technical support (KSC-2021-CRE-0025).
\end{acknowledgments}

\software{\texttt{Athena} \citep{stone08}, \texttt{VisIt} \citep{visit}}

\bibliographystyle{aasjournal}
\bibliography{mybib}

\appendix

\section{Seed Magnetic Fields}\label{app:seed-magnetic-fields}

In the \texttt{Athena} code, magnetic fields are face-centered and are updated by the edge-centered electromotive force ($\propto {\bf v} \times {\bf B}$) using the constrained transport algorithm.
Because our initial conditions have zero magnetic field at every \emph{active} face, the only way to inject magnetic fields into the domain is by having non-zero electromotive forces at the boundaries.
Even if $\mathbf{B}_\mathrm{in}$ is set parallel to $\mathbf{v}_\mathrm{in}$ for gas streams at the boundaries in our simulations, the fact that \texttt{Athena} defines the magnetic fields and velocity at the face centers and cell centers, respectively, enables $\mathbf{v}_\mathrm{in} \times \mathbf{B}_\mathrm{in}\neq 0$ at the domain boundaries, making the electromotive force nonzero and inducing seed magnetic fields in the active zones adjacent to the nozzles for an initial brief period of time.

\begin{figure}
    \epsscale{0.7}
    \plotone{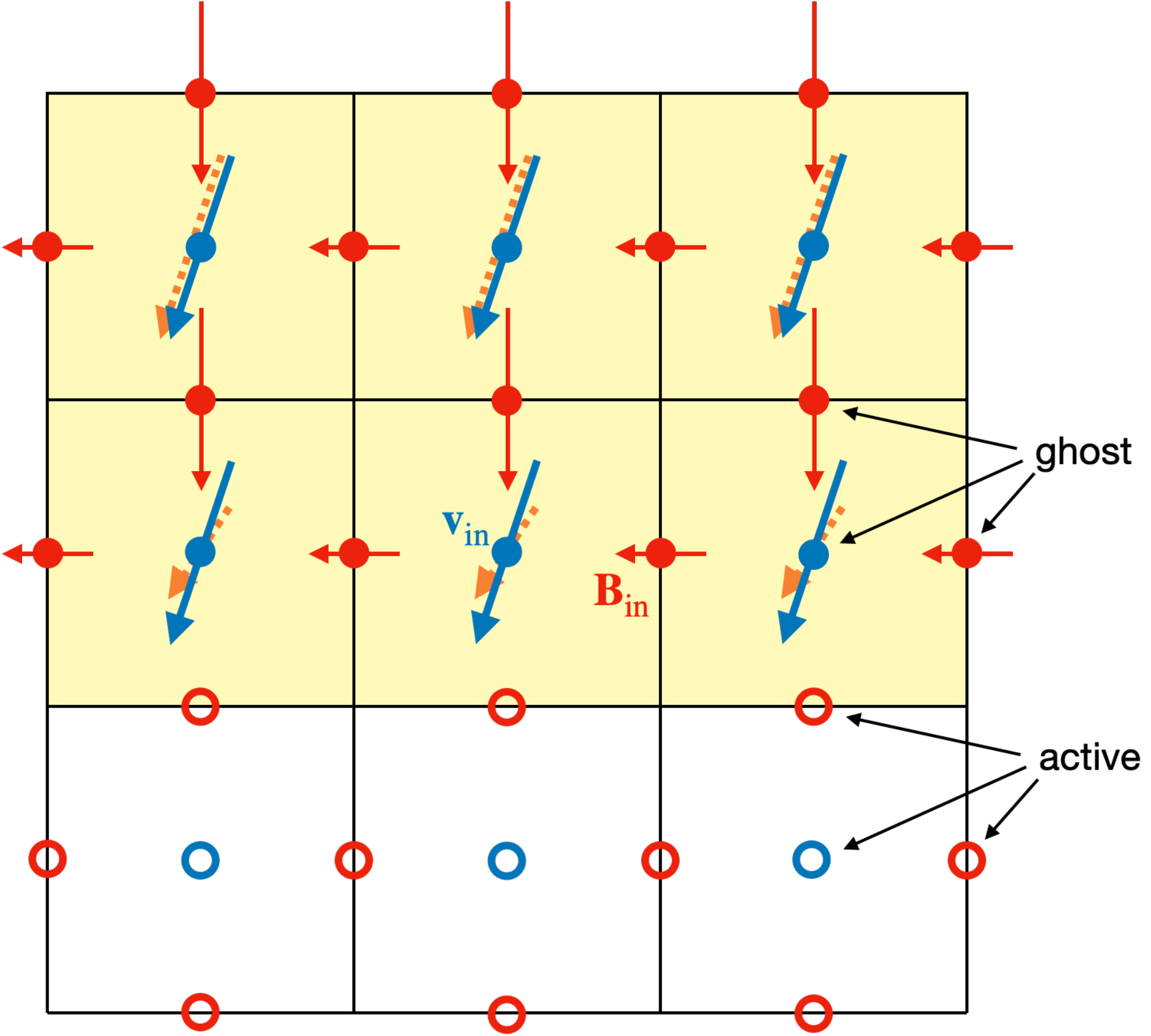}
    \caption{Illustration of the magnetic fields and velocity vectors in the ghost cells (shaded area) that belong to the nozzle at the positive $y$-boundary; the white area corresponds to the adjacent active cells. The blue arrows represent the velocity vectors, defined at cell centers. The solid and dashed arrows in red indicate the face- and cell-centered magnetic field vectors, respectively, with the latter computed from the former. Filled and open circles mark the ghost and active zones, respectively.
    See \autoref{app:seed-magnetic-fields} for details.}
    \label{fig:boundary-schematics}
\end{figure}

To illustrate this,  \cref{fig:boundary-schematics} diagrams a part of the computational domain near the positive-$y$ boundary.
The yellow shaded area indicates the ghost cells belonging to the upper nozzle and the white area marks the adjacent active cells.
The blue arrows at the cell centers represent $(v_x, v_y)$ of the stream at $t=0$, which are non-zero only in the ghost cells and zero in the active cells.
The red solid arrows at the cell faces indicate $(B_x, B_y)$ of the stream at $t=0$, which are also non-zero only in the ghost faces and zero in the active faces, while the red dashed arrows represent the cell-centered magnetic fields computed by averaging the neighboring face-centered fields. We note that the cell faces corresponding to the domain boundary are \emph{active} (i.e., $B_y$ at the border between the yellow and white areas is updated via \crefrange{eq:cont}{eq:Poisson}).
Because $B_y=0$ initially at those boundary faces, the cell-centered $B_y$ at the first ghost cells (adjacent to the active domain) is reduced by half, resulting in $\mathbf{B}_\mathrm{in}$ inclined to $\mathbf{v}_\mathrm{in}$, the latter of which is set to \cref{eq:vin}.
As a result, non-vanishing electromotive forces are assigned to the edges of the outermost active cells which subsequently induce magnetic fields into the computational domain.
We stress that this process occurs only for a few $\mathrm{Myr}$ in the very beginning: $B_y$ at the outermost \active faces soon attains the same values as in the ghost faces to satisfy \cref{eq:Bin}.
We also note that our initial conditions for the gas streams obey $\boldsymbol{\nabla}\cdot\mathbf{B}=0$ in the \emph{active} domain, which is preserved by the constrained transport algorithm in \texttt{Athena}.

\section{Magnetic Energy Conservation}\label{app:magnetic-energy-conservation}

Here we consider the role of advection into the domain in the growth of magnetic energy in our simulations.
We start with the equation for the rate of change of the total magnetic energy in the computational domain
\begin{equation}\label{eq:dEmagdt}
    \frac{dE_\mathrm{mag}}{dt} = \int \frac{\partial}{\partial t}\left(\frac{B^2}{8\pi}\right) dV
    = \frac{1}{4\pi} \int \mathbf{B}\cdot [\boldsymbol{\nabla}\times(\mathbf{v}\times\mathbf{B})] dV.
\end{equation}
Integrating \cref{eq:dEmagdt} by parts and applying the divergence theorem, one obtains
\begin{equation}\label{eq:magconserv-form1}
    \frac{dE_\mathrm{mag}}{dt} = \frac{1}{4\pi} \oint [(\mathbf{v}\times \mathbf{B})\times\mathbf{B}]\cdot {d\mathbf{A}} - \frac{1}{4\pi}\int \mathbf{v}\cdot\left[(\boldsymbol{\nabla}\times\mathbf{B})\times\mathbf{B}\right] dV,
\end{equation}
where $d\mathbf{A}$ denotes the area element.
The first term in the right hand side of \cref{eq:magconserv-form1} represents the Poynting flux integrated over the domain boundaries, while the second term is the amount of work done by the fluid against the Lorentz force per unit time.
It is evident that the first term vanishes when $\mathbf{v}\parallel \mathbf{B}$: there is no magnetic energy flux through the boundaries as long as the magnetic fields in the streams are parallel to the streaming velocity.
One can further expand the cross products in the first term to write
\begin{equation}\label{eq:magconserv-form2}
    \frac{dE_\mathrm{mag}}{dt} = -\oint \mathbf{v}\cdot\mathds{T}\cdot {d\mathbf{A}}  - \oint \frac{B^2}{8\pi}\mathbf{v}\cdot{d\mathbf{A}} - \frac{1}{4\pi}\int \mathbf{v}\cdot\left[(\boldsymbol{\nabla}\times\mathbf{B})\times\mathbf{B}\right] dV.
\end{equation}
In this form, the first and second term correspond to the work done by the Maxwell stress $\mathds{T} \equiv  B^2/(8\pi)\mathds{I} - \mathbf{B}\mathbf{B}/(4\pi)$ at the boundaries and the advection of magnetic energy by the inflowing gas, respectively.
Again, the two terms exactly cancel each other when $\mathbf{v}\parallel \mathbf{B}$.

As explained in  \autoref{app:seed-magnetic-fields}, $\mathbf{v}$ is not parallel to $\mathbf{B}$ at the domain boundaries for the initial $\sim10\,\mathrm{Myr}$, in which case the advection term is not offset by the Maxwell stress term, resulting in the growth of $E_\mathrm{mag}$.
One can estimate the maximum rate of the magnetic energy growth due to advection alone as
\begin{equation}
    \frac{dE_\mathrm{mag,adv}}{dt} = -\oint_\mathrm{nozzles} \frac{B_\mathrm{in}^2}{8\pi}\mathbf{v}_\mathrm{in}\cdot{d\mathbf{A}} = -\frac{2k_\mathrm{B}T_\mathrm{in}}{\beta_\mathrm{in}\mu_\mathrm{H}m_\mathrm{H}} \int_0^{\zeta_\mathrm{in}} \rho_\mathrm{in}\mathbf{v}_\mathrm{in}\cdot\hat{\mathbf{y}}\cos^2\left(\frac{\pi \zeta}{2\zeta_\mathrm{in}}\right) 2\pi \zeta d \zeta \approx 0.6 \frac{k_\mathrm{B}T_\mathrm{in}\dot{M}_\mathrm{in}}{\beta_\mathrm{in}\mu_\mathrm{H} m_\mathrm{H}},
\end{equation}
where \cref{eq:Bin-mag} is used.
For the parameters of model \texttt{B100}, $dE_\mathrm{mag,adv}/dt \sim 1.4\times 10^{49}\,\mathrm{erg\,Myr^{-1}}$.
This suggests that the total magnetic energy that would be advected (barring the work done by the Maxwell stress) into the computational domain  is $1.4\times 10^{50}\,\mathrm{erg}$ for initial $10\,\mathrm{Myr}$, which is a factor 3 smaller than the actual magnetic energy $E_\mathrm{mag} = 4.3\times 10^{50}\,\mathrm{erg}$ at $t=10\,\mathrm{Myr}$.
In contrast, the total magnetic energy advected into the computational domain would be $4.2\times 10^{51}\,\mathrm{erg}$ at the end of the run ($t=300\,\mathrm{Myr}$), which is about two orders of magnitude smaller than $E_\mathrm{mag}=7.2\times 10^{53}\,\mathrm{erg}$ at the same epoch.
Considering the work done by the Maxwell stress which tends to offset the magnetic energy growth by advection, the above result suggests that the actual magnetic energy advected through the nozzles should be negligible compared to what is generated by the last term in \cref{eq:magconserv-form2} via
an \ac{MHD} dynamo.
We conclude that while the inflow nozzles provide seed magnetic fields, it is growth via dynamo activity rather than advection into the domain that is responsible for level of the magnetic energy at late times.

\section{Mass Accretion Rates due to Maxwell and Reynolds Stresses}\label{app:accretion}

For gas to move radially inward while moving on an approximately circular orbit, it must lose angular momentum slowly.
Here we derive the theoretical accretion rates due to the Maxwell and Reynolds stresses.

Multiplying  the azimuthal component of \cref{eq:momentum} by $R$ yields
\begin{equation}\label{eq:angular-momentum-conservation}
    \frac{\partial(\rho R v_\phi)}{\partial t} + \boldsymbol{\nabla}\cdot \left(\rho R v_\phi\mathbf{v} + RP\mathbf{e}_\phi \right) =
    -R\mathbf{e}_\phi\cdot\left(\boldsymbol{\nabla}\cdot\mathds{T}\right) - 2R\Omega_p \rho v_R - \rho\frac{\partial\Phi_\mathrm{tot}}{\partial\phi},
\end{equation}
where $\mathbf{e}_\phi$ is the unit vector in the azimuthal direction.
The second and third terms in the left hand side of \cref{eq:angular-momentum-conservation} are the angular momentum flux due to macroscopic bulk fluid motion and the microscopic thermal motion of its constituent particles, respectively.
The three source terms in the right hand side are the torque density due to the Lorentz force, the Coriolis force, and gravity, respectively.
To focus on the radial mass inflow, we azimuthally average \cref{eq:angular-momentum-conservation}.
Integrating the resulting equation in the vertical direction assuming the flux through the vertical boundaries are negligible, one obtains
\begin{equation}\label{eq:angular-momentum-conservation-averaged}
    \frac{\partial\left<\rho R v_\phi\right>}{\partial t} + \frac{1}{R}\frac{\partial \left<R^2\rho v_R v_\phi\right>}{\partial R} = -\frac{1}{R}\frac{\partial \left<R^2T_{R\phi}\right>}{\partial R} - 2R\Omega_p\left<\rho v_R\right>,
\end{equation}
where $\left<X\right> \equiv (2\pi)^{-1}\iint X d\phi dz$ for any physical quantity $X$, and $T_{R\phi} = -B_R B_\phi / (4\pi)$ is the $R$--$\phi$ component of the Maxwell stress tensor.
Note $\left<\rho\partial\Phi_\mathrm{tot}/\partial \phi\right> = \left<\rho\partial\Phi_\mathrm{self}/\partial \phi\right> \approx 0$ unless there is a systematic azimuthal offset between the density and self-gravitational potential.
The magnetic torque term is due to the magnetic tension alone.

We decompose the velocity field into ordered and random components: $\mathbf{v} = v_\mathrm{circ}\mathbf{e}_\phi + \mathbf{u}$, where $v_\mathrm{circ} \equiv v_\mathrm{rot} - R\Omega_p$ is the circular velocity in the rotating reference frame and $\mathbf{u}$ is the random velocity.
Substituting $\mathbf{u}$ for $\mathbf{v}$ and using the continuity equation (\cref{eq:cont}), \cref{eq:angular-momentum-conservation-averaged} becomes
\begin{equation}\label{eq:perturbed-angular-momentum}
    \frac{\partial \left<\rho R u_\phi\right>}{\partial t} - \frac{\dot{M}_\mathrm{acc}}{2\pi R}\frac{\partial \left(Rv_\mathrm{circ}\right)}{\partial R} + \frac{1}{R}\frac{\partial  \left<R^2\rho u_R u_\phi\right>}{\partial R} = -\frac{1}{R}\frac{\partial \left<R^2T_{R\phi}\right>}{\partial R} - 2R\Omega_p\left<\rho u_R\right>,
\end{equation}
where $\dot{M}_\mathrm{acc} \equiv -2\pi R\left<\rho u_R\right>$ is the mass accretion rate at radius $R$.
Assuming a quasi-steady state (which turns out to be the case in our simulations) and neglecting the Coriolis term which is unimportant for small $R$, \cref{eq:perturbed-angular-momentum} is simplified to
\begin{equation}
\dot{M}_\mathrm{acc} \approx \dot{M}_\mathrm{M} + \dot{M}_\mathrm{R},
\end{equation}
where $\dot{M}_\mathrm{M}$ and $\dot{M}_\mathrm{R}$ are the mass accretion rates due to the Maxwell and Reynolds stress, defined in \cref{eq:appmdot}.
\Cref{fig:inflow_rate_t,fig:inflow_rate_R}
show that the mass accretion in our simulations is dominated by  $\dot{M}_\mathrm{M}$, that is, magnetic tension.

\end{document}